\documentclass[journal,10pt]{IEEEtran}

\usepackage{array,graphicx,cite,epsfig,amssymb,amsfonts}
\usepackage{bbm,amsmath,epstopdf,algorithm,algorithmic, multirow}
\usepackage{comment}
\usepackage{color}
\usepackage{url}
\usepackage{subfigure}
\usepackage{enumerate}

\newcommand{\be}{\begin{equation}}
\newcommand{\ee}{\end{equation}}
\def\bee#1\eee{\begin{align}#1\end{align}}
\newcommand{\bse}{\begin{subequations}}
\newcommand{\ese}{\end{subequations}}
\newcommand{\nnb}{\nonumber}

\newtheorem{theorem}{\textbf{Theorem}}
\newtheorem{proposition}{\textbf{Proposition}}
\newtheorem{lemma}{\textbf{Lemma}}

\newtheorem{example}{\textbf{Example}}

\newcommand{\specialcell}[2][c]{%
  \begin{tabular}[#1]{@{}c@{}}#2\end{tabular}}

\newcommand{\bm}[1]{\boldsymbol{#1}}

\ifodd 1
\else

\fi

\def \ISTR {}

\ifx \ISTR \undefined
\else
\fi

\ifx \ISTR \undefined
\else
\fi

\begin{document}

\ifx \ISTR \undefined
\title{Delay-Constrained Topology-Transparent Distributed Scheduling for MANETs}
\else
\title{Delay-Constrained Topology-Transparent Distributed Scheduling for MANETs \\ \vspace{0.5cm} \Large{Technical Report}  \vspace{0.5cm}}
\fi

\author{
Lei Deng$^*$,~\IEEEmembership{Member,~IEEE}, Fang Liu$^*$,
Yijin Zhang,~\IEEEmembership{Senior Member,~IEEE}, and Wing Shing Wong,~\IEEEmembership{Life Fellow,~IEEE}
\ifx \ISTR \undefined
\vspace{-1cm}
\else
\fi
\thanks{This work was supported in part by Schneider
Electric, Lenovo Group Ltd., China, and in part by the Hong Kong Innovation
and Technology Fund through the HKUST-MIT Research Alliance Consortium under Grant ITS/066/17FP,
in part by the Research Grants Council of the Hong Kong Special Administrative Region under Project GRF 14200217,
in part by NSFC under Grants 61902256 and 62071236, in part by
Tencent ``Rhinoceros Birds"-Scientific Research Foundation for Young Teachers of Shenzhen University,
and in part by the Fundamental Research Funds for the Central Universities of China under Grant 30920021127.
$^*$The first two authors contributed equally to the work. \emph{(Corresponding author: Lei Deng.)}}
\thanks{L. Deng is with College of Electronics and Information Engineering, Shenzhen University, Shenzhen 518060, China,
also with the Shenzhen Key Laboratory of Digital Creative Technology, Shenzhen 518060, China, and also with the Guangdong Province Engineering Laboratory for Digital Creative
Technology, Shenzhen 518060, China (e-mail: ldeng@szu.edu.cn).
Most of this work was done when L. Deng was with Department of Information Engineering, The Chinese University of Hong Kong, Hong Kong.}
\thanks{F. Liu and W. S. Wong are with Department of Information Engineering, The Chinese University of Hong Kong, Hong Kong (e-mail: \{lf015, wswong\}@ie.cuhk.edu.hk).}
\thanks{Y. Zhang is with School of Electronic and Optical Engineering, Nanjing University of Science and Technology, Nanjing 210094, China (e-mail: yijin.zhang@gmail.com).}
}

\maketitle

\begin{abstract}
Transparent topology is common in many mobile ad hoc networks (MANETs)
such as vehicle ad hoc networks (VANETs),  unmanned aerial vehicle (UAV) ad hoc networks, and wireless sensor networks
due to their  decentralization and mobility nature. There are many existing works on  distributed scheduling scheme design for
topology-transparent MANETs. Most of them focus on delay-unconstrained settings.
However, with the proliferation of real-time applications over wireless communications,
it becomes more and more important to support delay-constrained traffic in MANETs.
In such applications, each packet has a given hard deadline: if it is not delivered before its deadline, its
validity will expire and it will be removed from the system.
This feature is fundamentally different from the traditional
delay-unconstrained one. In this paper, we for the first time investigate distributed
scheduling schemes for a topology-transparent MANET to support delay-constrained traffic.
We analyze and compare probabilistic ALOHA scheme and deterministic sequence schemes,
including the conventional time division multiple access (TDMA), the Galois field (GF) sequence scheme proposed in \cite{chlamtac1994making},
and the combination sequence scheme that we propose for a special  type of sparse network topology.
We use both theoretical analysis and empirical simulations to compare all these schemes and
summarize the conditions under which different individual schemes perform best.
\end{abstract}

\section{Introduction}
An ad hoc network is \emph{topology-transparent} if the network topology is unknown to all network nodes.
Many mobile ad hoc networks (MANETs) have transparent topologies
since it is difficult or infeasible for individual nodes to acquire global
network connection information in real time, especially in the case of no centralized controller.
For example, network nodes in a vehicle ad hoc network (VANET) or an unmanned aerial vehicle (UAV) ad hoc network
move over time and thus the network topology changes over time; it is costly for
sensor nodes in a large-scale wireless sensor network to obtain
the whole network topology due to its large scale.
How to perform distributed scheduling to deliver packets under the topology-transparent setting has become a
vital research direction.

There are many solutions on this topic, including probabilistic schemes and deterministic schemes.
One conventional probabilistic scheme is
slotted ALOHA where each node transmits its packet at any slot with a common probability.
For deterministic scheme,  time division multiple access (TDMA),
where each node is assigned a unique slot to transmit, is a common option.
A variety of more sophisticated topology-transparent sequence schemes
have been proposed in the literature; see the survey paper \cite{kar2017survey} and the references therein.
Such schemes usually take into account the \emph{network density} $D$ (maximum number of interfering nodes among all nodes in the network).
They include algebraic approaches based on properties of Galois field (GF) \cite{chlamtac1994making, ju1998optimal},
combinatorial approaches  based on combinatorial structures like orthogonal arrays
and Steiner systems \cite{syrotiuk2003topology, colbourn2003steiner}, and number-theoretic approaches
based on Chinese remainder theorem \cite{su2015topology, su2016note}, etc.
Among them, the GF sequence scheme \cite{chlamtac1994making} is the most common one.


Most existing approaches for topology-transparent distributed scheduling
focus on delay-unconstrained traffic where a packet can be kept in the queue for however much time.
However, with the proliferation of real-time applications over wireless communications,
MANETs nowadays need to support more and more delay-constrained traffic.
Typical examples include multimedia wireless transmission system such as real-time
streaming and video conferencing via cellular or WiFi networks,
wireless cyber-physical systems (CPSs) such as  factory automation via wireless communications \cite{deng2017timely},
and networked control systems (NCSs) such as remote control (via wireless communications) of  UAVs \cite{baillieul2007control}.
In these  applications, each packet has a \emph{given} hard deadline: if it is not delivered before its deadline, it expires and  will be removed from the system. This feature is fundamentally different from the traditional
delay-unconstrained one.
There are many existing research works on delay-constrained wireless communications
where the major performance metric is \emph{timely throughput},
which is usually defined as the ratio of the number of packets that have been
delivered before expiration to the number of all generated packets \cite{hou2009qos,zhang2012combining,li2014optimal,deng2017timely}.
The concept of timely throughput is also
closely related to \emph{reliability}, which is a major performance metric
in ultra-reliable low latency communications (URLLC) in 5G \cite{bennis2018ultrareliable,singh2018contention,elayoubi2019radio,zhang2019achieving}.
However, those existing works only focus on networks with known (instead of transparent) topologies.

To the best of our knowledge, designing topology-transparent distributed scheduling schemes to support delay-constrained traffic in an MANET remains an open question.
In this work, this problem is investigated for the first time.
We use both theoretical analysis and empirical simulations to study probabilistic
ALOHA scheme and three deterministic sequence schemes under the delay-constrained setting.
The sequence schemes include \emph{TDMA} and \emph{the GF sequence scheme} \cite{chlamtac1994making} for general network density $D$,
and \emph{the combination sequence scheme} for a special  type of sparse network topology with $D=1$.
\ifx \ISTR \undefined
{Our main contributions of this work is to compare these schemes both theoretically and empirically
and summarize the conditions under which different individual schemes outperform others.}
\else
Our main contributions are listed as follows:

\begin{itemize}
\item We derive the exact average system timely throughput of TDMA
and theoretical lower bounds of the average system timely throughput of  ALOHA,
the GF sequence scheme and the combination sequence scheme;
\item {By leveraging existing results in a rather straightforward way,} we prove that ALOHA and the GF sequence scheme achieve better system performance than TDMA when~$D$ is small enough,
and prove that TDMA  has shorter sequence period than the GF sequence scheme when~$D$ is large enough;
\item We prove that when $D=1$ and $N \ge 10$ where $N$ is the number of transmitter-receiver pairs
in the network, the GF sequence scheme achieves better or equal system performance than TDMA, and when $D=N-1$ with $N$ a prime power,
TDMA achieves better or equal system performance than the GF sequence scheme;
\item We carry out extensive simulations to compare different schemes and summarize the conditions under which
different individual schemes outperform others.
\end{itemize}

\fi

\ifx \ISTR \undefined
\else
The rest of this paper is outlined as follows. We describe our system model and problem formulation in Sec.~\ref{sec:model}.
Then we analyze the probabilistic ALOHA scheme in Sec.~\ref{sec:aloha} and deterministic sequence schemes in Sec.~\ref{sec:sequence}.
We next compare different schemes via theoretical analysis in Sec.~\ref{sec:comparison} and via empirical simulations in Sec.~\ref{sec:simulation}.
Finally, we conclude this paper in Sec.~\ref{sec:conclusion}.
\fi

\ifx \ISTR \undefined
{Due to the page limit, we have moved most of technical proofs and some supplementary materials into our technical report \cite{TR}}.
\else
\fi

\section{System Model} \label{sec:model}

\textbf{Network Topology.} We consider an MANET with $N$ transmitters and $N$ receivers
(both indexed from 1 to $N$) which are geographically distributed
\ifx \ISTR \undefined
in an area.
\else
in an area as shown in Fig.~\ref{fig:model-network}.
\fi
A transmitter can transmit packets (or cause interference) to a receiver if their distance is less than or equal to $\Delta>0$,
which is called the communication range.
In our work, we assume that transmitter $i \in \{1,2,\cdots,N\}$ only needs to send information to receiver $i$; they form a pair, called pair $i$.
{
One practical example is a VANET where multiple vehicle-to-vehicle pairs
need to share data simultaneously \cite{wang2018privacy,liu2018intelligent,lakew2020routing,yue2019ai,paranthaman2019exploiting,
liu2019smart,wang2020topology}.
Another practical example of our model is that individual controllers send control messages
to their own UAVs via a shared wireless communication channel \cite{guo2019uav}.
In addition, our model can also be applied to D2D networks where multiple D2D pairs share the same wireless channel
to transmit data \cite{sun2019social,xu2016performance,zhang2020envisioning}.
}

In addition, transmitter $i$ causes interference to receiver $j \neq i$ if
their distance is within the communication range~$\Delta$.
In this case, we  call transmitter $i$  an \emph{interferer} of receiver $j$.
Otherwise, if their distance is larger than $\Delta$, transmitter~$i$ is not an interferer of receiver $j$.
{Our channel model is an \emph{unreliable collision channel}.
If both transmitter $i$ and any one or more interferers of receiver $i$ transmit a packet simultaneously,
collision happens and no packets of them can be delivered.
Even without collision, receiver $i$ can successfully receive a packet of transmitter $i$ with probability $p_i \in (0,1]$
if transmitter $i$ transmits a packet.
The successful probability~$p_i$ models the unreliability of wireless transmission due to shadowing and fading.
The successful probabilities could be different for different pairs (i.e., $p_i$ depends on $i$)
due to heterogeneous channel qualities.}
The network topology can change over time arbitrarily but satisfies the following two conditions:
\begin{enumerate}[(i)]
\item For any $i \in \{1,2,\cdots,N\}$, the distance between transmitter $i$ and receiver $i$ is always within $\Delta$;
\item At any time and at any location, there are no more than $D+1$ transmitters in any circle of radius $\Delta$,
where $D$ is a non-negative integer.
\end{enumerate}
Condition (i) shows that transmitter $i$ always establishes connection to
receiver $i$.
Condition (ii) is the density assumption of the network topology, which
means that any receiver can have at most $D$ interferers at any time excluding its own intended transmitter.
We also call $D$ the \emph{network density} of the network topology.  Note that $D$ is not necessarily equal to $N-1$. When the transmitters are distributed sparsely, $D$ can be far less than $N-1$.
In particular, when $D=0$, meaning that
all transmitters are distributed extremely sparsely, all receivers do not have any interferer
and thus all transmitters can transmit simultaneously without any collision.
To avoid such triviality, we assume that $D \ge 1$ in the rest of this paper.

The network topology is \emph{transparent} in the sense that all transmitters do not know the exact network topology but only the density $D$.
Furthermore, we assume that there is \emph{no feedback} from the receiver to the transmitter
about whether the transmitted packet is delivered successfully or not.

\ifx \ISTR \undefined
\else
\begin{figure}
    \centering
        \includegraphics[width=0.7\linewidth]{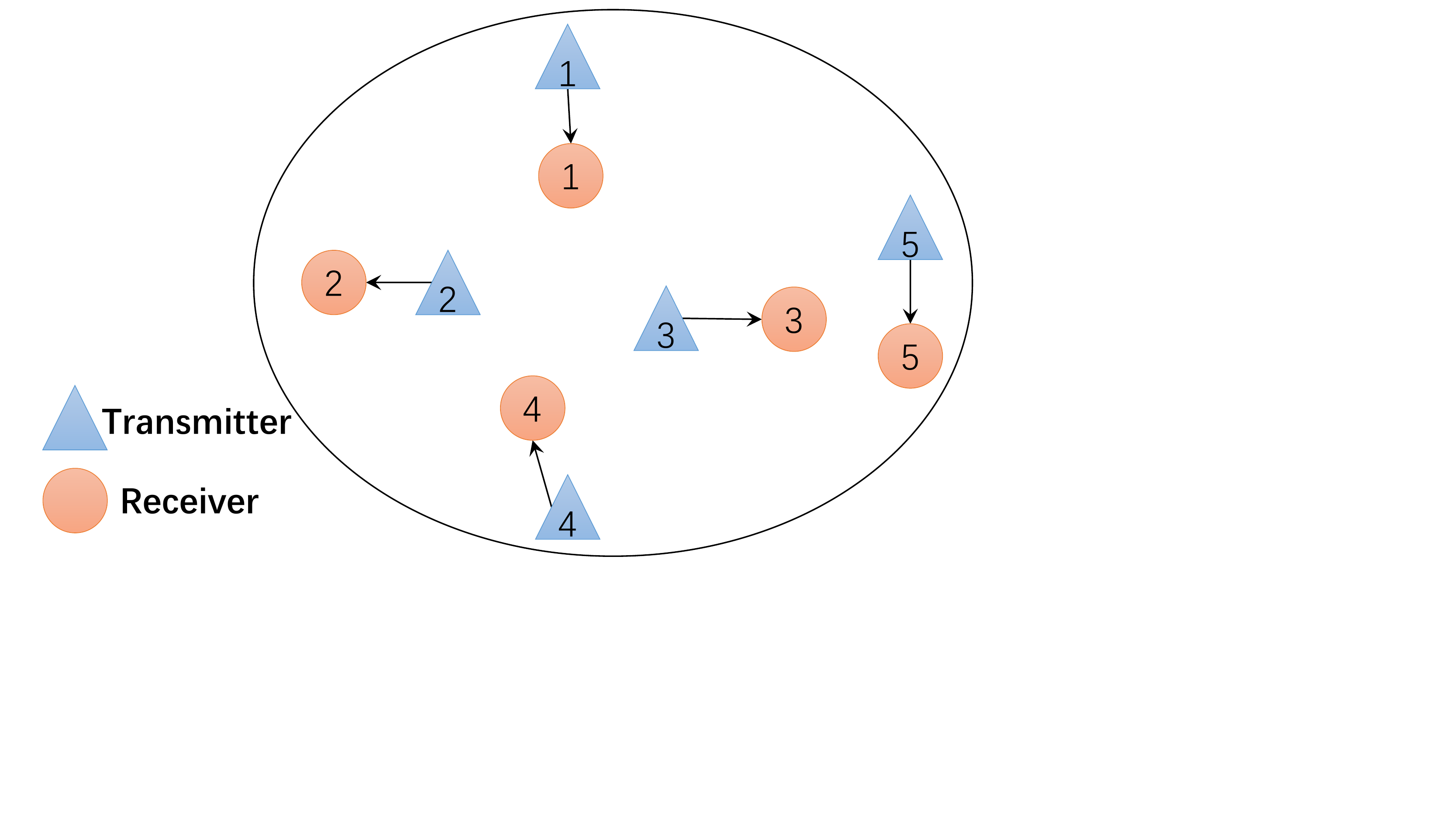}
    \caption{An example for the system model with $N=5$.} \label{fig:model-network}
\end{figure}
\fi


\textbf{Delay-Constrained Traffic Pattern.}
We consider a time-slotted system (indexed from slot 1) in which all nodes are time synchronized with no propagation delay.
We assume that the hard deadline of all packets in the system is $T$ slots, which
is specified by the application. In general,  the scheduling design and the system performance
are greatly influenced by the traffic pattern under the delay-constrained setting \cite{deng2017timely}.
In this work, as a first attempt to investigate the topology-transparent distributed scheduling under the delay-constrained setting,
we consider a simple yet common frame-synchronized traffic pattern \cite{hou2009qos,deng2017timely,deng2018delay},
which can find applications in CPSs \cite{kim2012cyber} and NCSs \cite{dengonstability2018}
where a system generates the control packets/messsages periodically.
\ifx \ISTR \undefined
Starting from slot 1,
\else
As shown in Fig.~\ref{fig:traffic}, starting from slot 1,
\fi
every~$T$ consecutive slots is called a frame, indexed from frame 1.
Therefore, frame $k$ consists of slot $(k-1)T+1$ to slot $kT$. We also call the application-specified hard deadline $T$ the \emph{frame length}.
Each of the $N$ transmitters generates a packet at the beginning of a frame,  which
will become expired and be removed from the system at the end of the frame.
Consider the example of UAVs. Each controller (transmitter) needs to send control messages to its controlled UAV periodically, and the period is $T$ slots. All the controllers' clocks are synchronized so that the starting time and the period are the same in all the controllers. This is an example for frame-synchronized traffic pattern. In addition, following \cite{hou2009qos,hou2010utility,deng2017timely}, we investigate the delay-constrained topology-transparent scheduling problem beginning with this special frame-synchronized traffic pattern.
\ifx \ISTR \undefined
In our technical report \cite{TR}, we also evaluate the performance of the proposed schemes under a poisson-arrival traffic pattern.
\else
Later in Sec.~\ref{subsec:sim-poisson}, we also evaluate the performance of the proposed schemes under a poisson-arrival traffic pattern.
\fi

The {timely throughput} of pair $i$ is defined as,
\be
R_i \triangleq \lim_{k \to \infty} \frac{ \mathbb{E} \left[  \substack{\text{number of pair-$i$ packets delivered before } \\ \text{expiration from slot 1 to slot $kT$}} \right]}{k},
\label{equ:def-R-i}
\ee
which only counts those packets that have been delivered before expiration \cite{hou2009qos,deng2017timely}.

Since there is one and only one new packet arrival in every frame,  \eqref{equ:def-R-i} implies that  the \emph{timely throughput} of pair $i$, i.e., $R_i$,
is the ratio of the expected number of packets that have been
delivered before expiration to the number of all generated packets of transmitter $i$.
Clearly, the maximum value of $R_i$ is 1.
In addition, the timely throughput defined in \eqref{equ:def-R-i} is the average probability that a pair-$i$ packet is delivered successfully before expiration.
Thus, it measures the reliability of pair $i$, which is a major performance metric in URLLC in 5G
\cite{bennis2018ultrareliable,singh2018contention,elayoubi2019radio,zhang2019achieving}.
Furthermore, we note that $R_i$ depends on the scheduling policy which will be explained next.

\ifx \ISTR \undefined
\else
\begin{figure}
  \centering
  \includegraphics[width=0.9\linewidth]{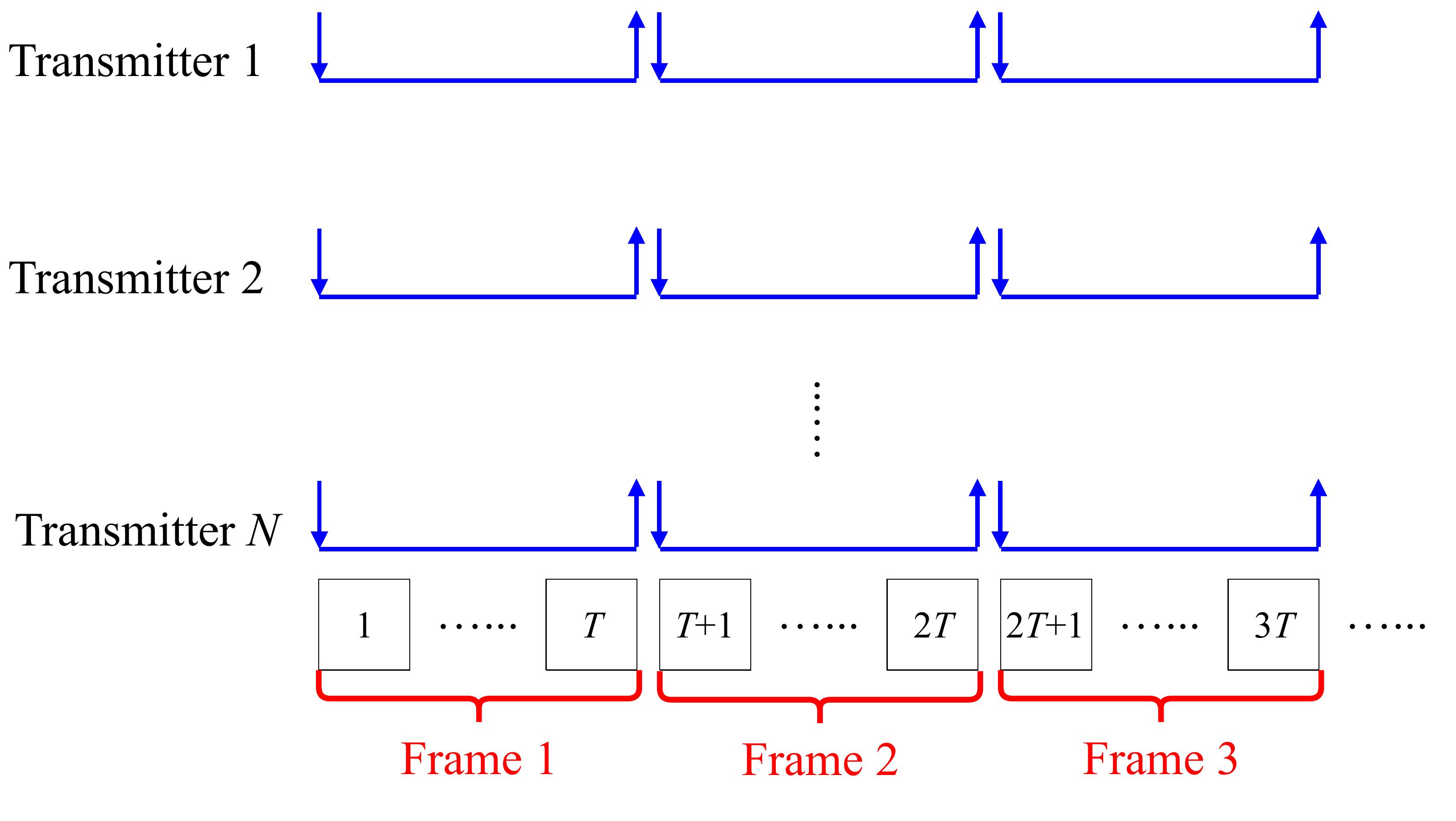}\\
  \caption{The frame-synchronized traffic pattern.}\label{fig:traffic}
\end{figure}
\fi

\textbf{Distributed Scheduling.}
Our goal is to design a distributed scheduling policy satisfying all aforementioned assumptions
to maximize the average system timely throughput, i.e.,
\be
\max_{\pi \in \Pi} \quad \frac{\sum_{i=1}^N R^{\pi}_i}{N},
\label{equ:problem-def}
\ee
where $\Pi$ is the set of all distributed scheduling policies and~$R^{\pi}_i$ is the achieved timely throughput of pair $i$ under policy~$\pi$.
The policy is distributed in the sense that each transmitter needs to determine its
own transmission strategy without the coordination of a centralized controller.

It is difficult to design the optimal distributed scheduling policy, i.e., solving \eqref{equ:problem-def} optimally.
In this work, we consider two popular types of distributed scheduling schemes: \emph{probabilistic ALOHA scheme} and
\emph{deterministic sequence schemes}.
\ifx \ISTR \undefined
{Since the probabilistic ALOHA scheme is relatively easy to design and analyze and due to the page limit, we move the details into out technical report \cite{TR}.
We use $\underline{R}^{\textsf{ALOHA}^*}(D,N,T)$ (see Equ. (6) in \cite{TR}) to denote the lower bound of the achieved average system timely throughput
under the probabilistic ALOHA scheme for given parameters $D$, $N$, and $T$.
In the following, we detail deterministic sequence schemes.}
\else
\fi


\ifx \ISTR \undefined
\else
{
\section{Probabilistic ALOHA Scheme} \label{sec:aloha}
We consider the conventional slotted ALOHA scheme \cite{roberts1975aloha}: in each slot, each transmitter transmits its packet with a probability $\delta \in (0,1]$.
Our goal is to design $\delta$ to maximize the average system timely throughput only based  on the value of $D$.
Note that each transmitter is not able to be aware of whether its transmission is successful or not due to the lack of feedback information. Then, in our ALOHA scheme, each transmitter insists to send a packet probabilistically until the deadline of this packet is expired. In this manner, every packet can be sent probabilistically in every slot of its deadline $T$ slots and can be delivered successfully if one transmission is successful.
Without loss of generality, we focus on the first frame from slot~1 to slot $T$.
Let $d_i(t)$ denote
the number of interferers of receiver $i$ in slot $t \in \{1,2,\cdots,T\}$, for $i \in \{1,2,\cdots,N\}$.
The probability that a packet of transmitter $i$ is delivered successfully in slot $t$ is
$\delta (1- \delta)^{d_i(t)}p_i$.
Then, a packet of transmitter $i$ is delivered successfully if it is delivered successfully at least once in the frame, and the probability of this event can be calculated as
\be
1-\prod_{t=1}^T \left[1- \delta(1-\delta)^{d_i(t)}p_i \right], \nnb
\ee
which decreases as $d_i(t)$ increases.
According to the density assumption in Sec.~\ref{sec:model}, i.e., $d_i(t) \le D, \forall i, t$,
we can get a lower bound of the timely throughput of pair $i$,
\be
\underline{R}^{\textsf{ALOHA}}_i(D,T,\delta)  = 1-\prod_{t=1}^T \left[1-\delta(1-\delta)^{D}p_i \right].
\label{equ:R-i-aloha}
\ee
It is straightforward to prove that $\underline{R}^{\textsf{ALOHA}}_i(D,T,\delta)$  in \eqref{equ:R-i-aloha} is maximized at
\be \label{equ:opt-q-aloha}
\delta^* = \frac{1}{D+1},
\ee
{which is the same as that in the delay-unconstrained slotted ALOHA with saturated traffic \cite[Chapter 5.3.2]{kurose2013computer}
where all $D$ stations  transmit/retransmit their packets with probability $p_i$ and each station always has a new packet arrival once its packet has been delivered successfully.}

The corresponding lower bound of pair-$i$ timely throughput is
\be
\resizebox{.892\linewidth}{!}{$\underline{R}_i^{\textsf{ALOHA}^*}(D,T) = 1-\prod_{t=1}^T \left[1- \frac{1}{D+1} \left(1- \frac{1}{D+1}\right)^{D}p_i \right]$},
\ee
and the corresponding lower bound of the average system timely throughput is
\bee
& \underline{R}^{\textsf{ALOHA}^*}(D,N,T) = \frac{\sum_{i=1}^N \underline{R}_i^{\textsf{ALOHA}^*}(D,T)}{N}.
\label{equ:avg-system-throughput-ALOHA}
\eee

}
\fi

\section{Deterministic Sequence Schemes} \label{sec:sequence}
In sequence schemes, we pre-assign any transmitter $i$ a binary sequence $\bm{S}_i=(S_i(1), S_i(2), \cdots)$
with the convention that $S_i(t)=1$ means that transmitter $i$ will transmit its packet at slot $t$ and $S_i(t)=0$ means that
it will remain idle at slot $t$. Thus, following the assigned sequence,
each transmitter will either transmit or not in any slot. The sequence schemes are distributed in the sense that there is no need to involve a centralized controller once the sequences are assigned to transmitters.
They are deterministic schemes in contrast to the probabilistic ALOHA scheme.
{We remark that in the sequence-based schemes, a preliminary is to perform sequence allocation.
A common solution is to pre-assign sequences for users, which needs to know the total number of pairs, i.e., $N$, in advance.
We use this approach in our paper. For example, consider $N$ pairs of UAVs and controllers. Before they perform task by forming an MANET, we pre-assign each pair a sequence according to our sequence scheme.
Once the sequences are assigned to pairs, each transmitter can work distributedly according to its assigned sequence.
A more practical solution is to automatically allocate sequences relying on some extra knowledge. For example, reference \cite{wong2014transmission} describes a method
that a user can automatically get a sequence based on its geographic location.
Reference \cite{Wu2014Safety} introduces an allocation method for VANET with the help of roadside nodes or roadside units near highway entrances or toll booths.}

If in slot $t$, $S_i(t)=1$ and $S_j(t)=0$ for any interferer $j$ of receiver $i$, we call such a $t$ a \emph{collision-free slot} of transmitter $i$ (or sequence $\bm{S}_i$).
A packet of transmitter~$i$ can be delivered successfully with probability $p_i$ in those collision-free slots and
no successful delivery happens in other slots.
Note that for a given sequence scheme,
whether slot $t$ is a collision-free slot of transmitter~$i$ depends on the network topology.
Since the traffic pattern is fixed, the pair $i$'s timely throughput is determined by the set of all collision-free slots of transmitter~$i$.

In general, the sequence could be in an arbitrary form and of an infinite-dimension design space.
However, in our work, due to the periodical nature of the traffic pattern,
we only consider \emph{periodic} sequences in order to simplify the design. Specifically, a periodic sequence $\bm{S}=(S(1), S(2), \cdots)$ with period $L$  satisfies $S(t) = S(t-L), \forall t > L$,
i.e., $\bm{S}=(S(1), S(2), \cdots, S(L), S(1), S(2), \cdots, S(L), \cdots)$.
Thus, a periodic sequence with period $L$ is completely determined by its first $L$ elements.
We then represent a periodic sequence with period $L$ by
a sequence of finite length $L$, i.e., $\bm{S}=(S(1), S(2), \cdots, S(L))$.
For a sequence of period $L$, starting from the first period, every $T$ periods is called a super period (of in total $TL$ slots),
indexed from super period 1.
Clearly, super period $k$ is from period $(k-1)T+1$ to period $kT$. We establish the following result.
\begin{theorem} \label{thm:theory-R-i-L-T}
If the sequence of transmitter $i$ is of period $L$ and
the set of its collision-free slots in any super period $k \in \{1,2,\cdots\}$ is $\{t_k+ (k-1)TL + mL: m=0,1,\cdots,T-1,1\leq t_k \leq L\}$, then
the following results hold.
\begin{itemize}
\item Case 1: If  $L \ge T$,
 the timely throughput of pair $i$ is
\be
R_i^{\textsf{Case-1}}(L,T) =  \frac{T}{L} \cdot p_i.
\label{equ:avg-1}
\ee
\item Case 2: If $L < T$,
 the timely throughput of pair $i$ is
\bee
 \resizebox{.892\linewidth}{!}{$R_i^\textsf{Case-2}(L,T)  = \frac{  \alpha \left[1-(1-p_{i})^{ \left \lceil \frac{T}{L} \right \rceil} \right]+\beta \left[1-(1-p_{i})^{ \left \lfloor \frac{T}{L} \right \rfloor} \right] }{L}$},
\label{equ:avg-2}
\eee
where $ \alpha= (T \bmod L)$ and $\beta=L-\alpha$.
\end{itemize}
\end{theorem}
\ifx \ISTR \undefined
\else
\begin{IEEEproof}
Please see Appendix~\ref{app:proof-of-lemma-theory-R-i-L-T}.
\end{IEEEproof}
\fi

The condition for the sequence in Theorem~\ref{thm:theory-R-i-L-T} means that
there is  \emph{exactly} one collision-free slot in any period of any super period and its location has the same offset relative to the beginning of the period
but the offset, i.e., $t_k$, could be different for different super periods. For simplicity, we call it \emph{location-fixed condition}.
Theorem~\ref{thm:theory-R-i-L-T} shows that if a sequence satisfies the location-fixed condition,
we can use \eqref{equ:avg-1} and \eqref{equ:avg-2} to obtain the exact timely throughput.
In addition, if the set of collision-free slots of a sequence in any super frame $k$ includes
some extra slots in addition to $\{t_k+ (k-1)TL + mL: m=0,1,\cdots,L-1\}$, we can use \eqref{equ:avg-1} and \eqref{equ:avg-2} to obtain a lower bound of the timely throughput.

In the special case of $p_i=1$, every packet of transmitter $i$ will be delivered successfully with certainty
if no interferer of receiver $i$ transmits simultaneously. This special case is called the \emph{perfect-channel case}.
It is straightforward to see that when $p_i=1$, \eqref{equ:avg-2} becomes
$
R_i^{\textsf{Case-2}}(L,T)=1.
$
This means that pair $i$ achieves its maximum value 1 where the sequence period $L$ is not greater than the frame length $T$.
Therefore, one direction to find best sequences in the perfect-channel case
is to find a sequence set of period $L$ such that each one has (at least) one collision-free-slot in a period subject to
the topology density constraint $D$. In addition, we should try to minimize the sequence period $L$ such that $L \le T$.
For the imperfect-channel case, i.e., $p_i \in (0,1)$, we also provide
a reason to minimize the sequence period $L$.

\begin{lemma} \label{lem:R-1-R-2-decrease-with-L}
$R_i^{\textsf{case-1}}(L,T)$ in \eqref{equ:avg-1}  strictly decreases as $L$ increases.
When $p_i=1$, $R_i^{\textsf{case-2}}(L,T)$ in \eqref{equ:avg-2}  remains to be constant 1 for all $L \le T$.
When $p_i \in (0,1)$, $R_i^{\textsf{case-2}}(L,T)$ in \eqref{equ:avg-2} strictly decreases as $L$ increases.
\end{lemma}
\ifx \ISTR \undefined
\else
\begin{IEEEproof}
Please see Appendix~\ref{app:proof-of-lem-R-1-R-2-decrease-with-L}.
\end{IEEEproof}
\fi

Lemma~\ref{lem:R-1-R-2-decrease-with-L} shows that if we can find a sequence set assigned to $N$ transmitters
such that each sequence has one collision-free slot in a period and satisfies the location-fixed condition,
we should try to minimize the sequence period $L$ to increase the  average system timely throughput.

We will next introduce three types of sequence schemes.
The first one is the conventional TDMA scheme,
which guarantees that each transmitter/sequence has exactly one collision-free slot in a period for any network topology.
The second one is the GF topology-transparent scheduling sequence scheme proposed in \cite{chlamtac1994making}.
For simplicity, we call it \emph{the GF sequence scheme}.
It guarantees at least one collision-free slot for each transmitter/sequence in a period
for any network topology with density $D$.
The last one is called \emph{the combination sequence scheme} designed  for the special case of $D=1$.
It is ``optimal" in the sense that it finds the minimal sequence period $L$ such that
each sequence has at least one collision-free slot in a period for any network topology with density~$D=1$.

\subsection{TDMA} \label{subsec:TDMA}
The simplest sequence scheme is the conventional TDMA scheme where we assign the transmission token in a round-robin manner.
Specifically, the sequence period is $L=N$ and the sequence for transmitter $i$, i.e., $\bm{S}_i$, satisfies
\be
S_i(t) =
\left\{
  \begin{array}{ll}
    1, & \hbox{if $t=i$;} \\
    0, & \hbox{otherwise.}
  \end{array}
\right.
\forall t=1,2,\cdots, N.
\ee

\ifx \ISTR \undefined
\else
For example, when $N=3$, the sequence set is
\be
\bm{S}_1 =(1,0,0), \bm{S}_2 =(0,1,0), \bm{S}_3 =(0,0,1). \nnb
\ee
\fi

The TDMA scheme guarantees that each transmitter has
exactly one collision-free slot within a period $L=N$ for any network topology
and any sequence $\bm{S}_i$ satisfies the location-fixed condition.
Then according to Theorem~\ref{thm:theory-R-i-L-T},
if $ N\geq T $,  the timely throughput of transmitter $i$
is $R_i^{\textsf{case-1}}(N,T)$ and the average system timely throughput is
\be
R^{\textsf{TDMA-1}}(N,T)= \frac{\sum_{i=1}^N R_i^{\textsf{case-1}}(N,T)}{N}.
\label{equ:avg-R-TDMA-1}
\ee
If $ N<T $, the timely throughput of transmitter $i$ is $R_i^{\textsf{case-2}}(N,T)$ and
the average system timely throughput is
\bee
&R^{\textsf{TDMA-2}}(N,T) =  \frac{\sum_{i=1}^N R_i^{\textsf{case-2}}(N,T)}{N}.
\label{equ:avg-R-TDMA-2}
\eee

Note that in TDMA, we guarantee that a sequence will not be blocked\footnote{A sequence $\bm{S}$ is blocked by sequences $\bm{S}_1,\bm{S}_2,\cdots,\bm{S}_k$
if there does not exist a slot $t$ such that $S(t)=1$ while $S_i(t)=0, \forall i=1,2,\cdots,k$.} by all other $N-1$ sequences,
regardless of the network topology. Thus, the results
in \eqref{equ:avg-R-TDMA-1} and \eqref{equ:avg-R-TDMA-2} hold for any
network topology with any network density $D$ which could change in any slot.

\subsection{The GF Sequence Scheme} \label{subsec:GF}
Different from TDMA, the GF sequence scheme proposed in \cite{chlamtac1994making}
can exploit the sparsity of the network topology, which guarantees that any sequence will not be blocked by any other $D$ sequences.
In the GF sequence scheme, a sequence period consists of $q$ sub-periods each of which is of length $ q $, where $q$ is a prime power.
Thus, the sequence period  is $L=q^2$.
The construction of a sequence is as follows.
Each transmitter $i\in \{1,2,\ldots, N \}$ is assigned a sequence according to
a unique polynomial $ f_i(e) $ of degree at most $k$ (where $k$ is a nonnegative integer)
over Galois field $GF(q)$.
We index the elements of $GF(q)$ from 1 to $q$ and the $x$-th element is denoted by $e_x$.
The value $ f_i(e_x) $ determines the transmission slot for transmitter~$i$ in the $ x $-th sub-period in the following manner: if the value of $f_i(e_x)=e_y$, i.e., the $y$-th element in $GF(q)$, we set the $y$-th slot in the $x$-th sub-period to be 1 and set other slots to be 0.

Within a period, the number of 1s of each transmitter is  $q$ since each sub-period contains exactly one 1. The number of conflicting 1s for any two transmitters within a period is at most $ k $ due to the following fact: for any two polynomials of degree at most $ k $ over $GF(q)$, their difference has at most $ k $ roots.
Thus as long as $ q-kD\geq 1 $, any sequence will not be blocked by any other $D$ sequences
and thus any transmitter can be guaranteed to have at least one collision-free slot.

The total number of polynomials with degree at most $k$ over $GF(q)$ is $ q^{k+1} $.
Then as long as $q^{k+1} \ge N$, we can guarantee that each transmitter can get a unique polynomial
and thus a unique sequence.
Therefore, to minimize the sequence period $L=q^2$, for given $N$ and $D$, we need to find the smallest prime power $q$ to satisfy
\be
\left\{
  \begin{array}{ll}
    q-kD\geq 1,  \\
    q^{k+1} \geq N.
  \end{array}
\right.
\label{equ:min-q-GF}
\ee
The smallest prime power $q$ satisfying \eqref{equ:min-q-GF} is denoted by~$q(D,N)$ and the corresponding sequence period is~$L=q^2(D,N)$.
\ifx \ISTR \undefined
{We give an example for GF sequence in our technical report \cite{TR}.}
\else
\begin{example}
Given $ N=4 $ and $D=1$,  we can find that $ q=2 $ is the smallest prime power that satisfies \eqref{equ:min-q-GF} where we set $k=1$ correspondingly. Thus $q(1,4)=2$.
For the 4 transmitters, we choose 4 polynomials with degree at most $ 1 $ over Galois field $ GF(2)=\{0,1 \} $ as follows:
\be
f_1(x)=0, f_2(x)=1, f_3(x)=x, f_4(x)=1+x.
\ee
In the first sub-period, $ f_1(0)=0, f_2(0)=1, f_3(0)=0, f_4(0)=1 $. In the second sub-period, $ f_1(1)=0, f_2(1)=1, f_3(1)=1, f_4(1)=0 $. Thus the sequence set is as follows,
\[
\resizebox{\linewidth}{!}{$\bm{S}_{1}=(1,0,1,0), \bm{S}_{2}=(0,1,0,1),\bm{S}_{3}=(1,0,0,1), \bm{S}_{4}=(0,1,1,0).$}
\]
We can check that each sequence has at least one collision-free slot in the case of $D=1$
since each sequence will not be blocked  by any other sequence.
\end{example}
\fi

When the network topology changes slowly in the sense that the topology is fixed in any super period,
any  GF sequence  has \emph{at least} one collision-free slot in any period and the offsets are the same in all periods in a super period.
Thus, the GF sequence scheme satisfies the location-fixed location possibly with some extra collision-free slots in the slowly-changing topology scenario.
Then according to Theorem~\ref{thm:theory-R-i-L-T},
the timely throughput of pair $i$ is lower bounded by $R_i^{\textsf{case-1}}(q^2(D,N),T)$
when $q^2(D,N) \ge T$ and is lower bounded by $R_i^{\textsf{case-2}}(q^2(D,N),T)$ when $q^2(D,N) < T$.
Thus, if the network topology does not change in any super period,
in the case of  $q^2(D,N) \ge T$, the average system timely throughput is lower bounded by
\be
\underline{R}^{\textsf{GF-1}}(D, N, T) =  \frac{\sum_{i=1}^N R_i^{\textsf{case-1}}(q^2(D,N),T)}{N};
\label{equ:avg-R-GF-1}
\ee
in the case of $q^2(D,N) < T$, the average system timely throughput is lower bounded by
\bee
& \underline{R}^{\textsf{GF-2}}(D, N, T) =  \frac{\sum_{i=1}^N R_i^{\textsf{case-2}}(q^2(D,N),T)}{N}.
\label{equ:avg-R-GF-2}
\eee

\subsection{The ``Optimal" Combination Sequence Scheme for $D=1$} \label{subsec:combination}
In the sequence scheme design, there is an interesting and important
combinatorial problem: what is the minimum length of the sequences
such that there exists a set of at least $N$ sequences each of which has at least one collision-free slot
for any network topology with density $D$?
We denote such minimum length by $L^{\min}(D,N)$.
\ifx \ISTR \undefined
{For $D=1$, we can explicitly characterize $L^{\min} (1,N)$.
We construct the sequence set according to all combinations of $L^{\min} (1,N)$ choosing $\left\lceil \frac{L^{\min} (1,N)}{2} \right\rceil$,
and then arbitrarily select $N$ sequences and assign them to $N$ pairs. This is the combination sequence scheme.
Due to the page limit, we move  details of the combination sequence scheme to our technical report \cite{TR}.
}
\else
For $D=1$, we have the following result.

\begin{proposition} \label{prop:min-L-for-D=1}
For any $N$, we have
\be
L^{\min} (1,N) = \min \left \{L \in \mathbb{Z}^+: \binom{L}{\left\lceil \frac{L}{2} \right\rceil} \ge N \right \}.
\ee
\end{proposition}
\begin{IEEEproof}
Please see Appendix~\ref{proof-of-thm-min-L-for-D=1}.
\end{IEEEproof}

Clearly we have $L^{\min}(1,N) \in \{1,2,\cdots,N\}$ and $\binom{L}{\left\lceil \frac{L}{2} \right\rceil}$ increases with respect to $L$.
Thus, we can use an efficient binary-search scheme in the range $\{1,2,\cdots,N\}$ to find $L^{1,\min}(N)$.

\begin{example}
Consider $N=10$. We have $L^{\min} (1,N)=5$.
Then we construct the sequence set according to all combinations of $L^{\min} (1,N)$ choosing $\left\lceil \frac{L^{\min} (1,N)}{2} \right\rceil$,
and then we select the first $N$ sequences and assign them to $N$ pairs.
We have the following $\binom{L^{\min} (1,N)}{\left\lceil \frac{L^{\min} (1,N)}{2} \right\rceil} = \binom{5}{3} = 10$ sequences,
\bee
& \bm{S}_1=(00111), \bm{S}_2=(01011), \bm{S}_3=(01101), \bm{S}_4=(01110), \nnb \\
& \bm{S}_5=(10011), \bm{S}_6=(10101), \bm{S}_7=(10110), \bm{S}_8=(11001), \nnb \\
& \bm{S}_9=(11010), \bm{S}_{10}=(11100). \nnb
\eee
Here $\bm{S}_1=(00111)$ means that the combination is to choose the last three slots from the total five slots.
All other sequences follow the similar rule. We can check that each sequence will not be blocked by any other sequence.
\end{example}

The combination sequence scheme also satisfies the location-fixed location possibly with some extra collision-free slots
in the slowly-changing topology scenario. Thus, according to Theorem~\ref{thm:theory-R-i-L-T},
if the network topology does not change in any super period, in the case of $L^{\min}(1,N) \geq T$, the average system timely throughput is lower bounded by
\bee
\small
&  \underline{R}^{\textsf{Combination-1}}(L^{\min}(1,N), T)  \nnb \\
& = \frac{\sum_{i=1}^N R_i^{\textsf{case-1}}(L^{\min}(1,N),T) }{N} = \frac{T\sum_{i=1}^{N} p_i}{N L^{\min}(1,N)};
\label{equ:avg-R-combination-1}
\eee
in the case of $L^{\min}(1,N) < T$, the average system timely throughput is lower bounded by
\bee
&\resizebox{.892\linewidth}{!}{$R^{\textsf{Combination-2}}(L^{\min}(1,N),T) = \frac{\sum_{i=1}^N R_i^{\textsf{case-2}}(L^{\min}(1,N),T) }{N}$}\nnb \\
&
\resizebox{.892\linewidth}{!}{$  =  \frac{\sum\limits_{i=1}^{N} \left(\alpha \left[1-(1-p_{i})^{ \left \lceil \frac{T}{L^{\min}(1,N)} \right \rceil} \right]+\beta \left[1-(1-p_{i})^{ \left \lfloor \frac{T}{L^{\min}(1,N)} \right \rfloor} \right] \right)}{NL^{\min}(1,N)},$}
\label{equ:avg-R-combination-2}
\eee
where $\alpha = (T \bmod L^{\min}(1,N))$ and $\beta = L^{\min}(1,N)- \alpha$.

\fi

\vspace{-0.2cm}

\section{Theoretical Comparison} \label{sec:comparison}
\ifx \ISTR \undefined
\else
In Sec.~\ref{sec:sequence}, we presented exact value or lower bounds of average system timely throughput
for different schemes.
Based on these results, we compare ALOHA and TDMA in Sec.~\ref{subsec:compare-aloha-TDMA} in terms of average system timely throughput.
According to Lemma~\ref{lem:R-1-R-2-decrease-with-L}, the sequence period is a key performance metric to the system,
we thus compare the sequence periods of all sequences schemes. Particularly,
we  compare the GF sequence scheme and TDMA in Sec.~\ref{subsec:compare-GF-and-TDMA} for general~$D$
and compare  TDMA, the GF sequence scheme, and the combination sequence scheme in Sec.~\ref{subsec:compare-seqs-for-D=1} for the special case of $D=1$.
In the following discussions, we assume
that the number of pairs $N$ and the frame length $T$ are given.
\fi

\subsection{Comparison between ALOHA and TDMA} \label{subsec:compare-aloha-TDMA}
First, we analyze the lower bound of the average system timely throughput of ALOHA.
\begin{theorem} \label{thm:aloha-decreasing-with-D}\
\ifx \ISTR \undefined
$\underline{R}^{\textsf{ALOHA}^*}(D,N,T)$ is strictly decreasing with respect to $D$.
\else
$\underline{R}^{\textsf{ALOHA}^*}(D,N,T)$ in \eqref{equ:avg-system-throughput-ALOHA} is strictly decreasing with respect to $D$.
\fi
\end{theorem}
\ifx \ISTR \undefined
\else
\begin{IEEEproof}
Please see Appendix~\ref{app:proof-of-thm-aloha-decreasing-with-D}.
\end{IEEEproof}
\fi

To compare ALOHA and TDMA, recall that the average  system timely throughput of TDMA does not vary with respect to~$D$.
Thus, we can find a smallest density~$D$ (denoted as $D^*$) such that
$\underline{R}^{\textsf{ALOHA}^*}(D,N,T) \ge  R^{\textsf{TDMA-1}}(N,T)$
when $N \ge T$ or
$\underline{R}^{\textsf{ALOHA}^*}(D,N,T) \ge  R^{\textsf{TDMA-2}}(N,T)$
when $N < T$.
Therefore, according to Theorem~\ref{thm:aloha-decreasing-with-D},
it follows that ALOHA has larger average system timely throughput than TDMA when $D \le D^*$.
In addition, such a $D^*$ can be found by an efficient binary-search scheme. We take the convention that  $D^*=-\infty$ if we cannot find such a $D$,
which means that TDMA is better than the lower bound of ALOHA for all $D \ge 1$.

\vspace{-0.2cm}

\subsection{Comparison between the GF Sequence Scheme and TDMA} \label{subsec:compare-GF-and-TDMA}
In the GF sequence scheme, the sequence period is $ L=q^2(D,N) $. We first establish the following result.

\begin{lemma} \label{lem:q-non-decreasing}
In the  GF sequence scheme, $L=q^2(D,N)$ is non-decreasing with respect to $D$.
\end{lemma}
\begin{IEEEproof}
For each $D$, we find the smallest prime power~$q$ satisfying \eqref{equ:min-q-GF}, i.e., $q(D,N)$. The result follows from the fact that the prime power $q$ satisfying \eqref{equ:min-q-GF} with density $D$ also satisfies \eqref{equ:min-q-GF} with any density $D' < D$.
\end{IEEEproof}

By combining Lemma~\ref{lem:R-1-R-2-decrease-with-L} and Lemma~\ref{lem:q-non-decreasing}, it follows that
the lower bound of average system timely throughput of the GF sequence scheme is non-increasing as~$D$ increases.
Therefore, similar to Sec.~\ref{subsec:compare-aloha-TDMA}, there exists a~$D^*$ such that
the GF sequence scheme has better average  system timely throughput than TDMA when $D \le D^*$.
On the other hand, we show that when $D$ is large enough, the sequence period of TDMA is less than that of the GF sequence scheme.

\begin{proposition} \label{prop:D-large-than-sqrt-2N-TDMA-better-than-GF}
When $D > \sqrt{2N}$, the sequence period of TDMA is less than that of the GF sequence scheme, i.e., $N < q^2(D,N)$.
\end{proposition}
\begin{IEEEproof}
Note that the sequence set in the GF sequence scheme is a ZFD code (see the definition in \cite{gyHori2008coding})
and thus the result follows from \cite[Theorem 2]{gyHori2008coding}.
\end{IEEEproof}

\ifx \ISTR \undefined
{In our technical report, we give an example to illustrate Proposition~\ref{prop:D-large-than-sqrt-2N-TDMA-better-than-GF}.}
\else
\begin{example}
For $N=100$, in Table~\ref{tab:q-D-100},
we list $q(D,N)$ for different $D$ and compare the sequence period of the GF sequence scheme which is $q^2(D,100)$
and the sequence period of TDMA which is $N$. In this case, the critical density is $D^*=4$.
In addition, when $D > \sqrt{2N} = \sqrt{200} = 14.1$, we can see that the sequence period of TDMA, i.e., $N$, is smaller
than that of the GF sequence scheme.
\end{example}

\begin{table*}[t]
\centering
\caption{$q(D,100)$ with different $D$. \label{tab:q-D-100}}
\begin{tabular}{|l|l|l|l|l|l|l|l|l|l|l|l|l|l|l|l|l|}
\hline
$D$                 & 1   & 2   & 3   & 4   & 5   & 6   & 7   & 8   & 9   & 10  & 11  & 12  & 13  & 14  & 15 & 16  \\ \hline
$q(D,100)$          & 4   & 5   & 7   & 9   & 11  & 11  & 11  & 11  & 11  & 11  & 13  & 13  & 16  & 16  & 16 & 17 \\ \hline
$q^2(D,100) $       & 16  & 25  & 49  & 81  & 121 & 121 & 121 & 121 & 121 & 121 & 169 & 169 & 256 & 256 & 256 & 289\\ \hline
$q^2(D,100) < 100?$ & Yes & Yes & Yes & Yes & No  & No  & No  & No  & No  & No  & No  & No  & No  & No  & No  & No \\ \hline
\end{tabular}
\end{table*}
\fi

Proposition \ref{prop:D-large-than-sqrt-2N-TDMA-better-than-GF}  shows that
the average system timely throughput of TDMA is larger than the lower bound
of the average system timely throughput of the GF sequence scheme when $ D>\sqrt{2N} $. However,
it does not mean that the actual average system timely throughput
of TDMA is larger than that of the GF sequence scheme when $ D>\sqrt{2N} $.
We will compare their actual performance by simulations in Sec.~\ref{sec:simulation}.
However in the special case of $D=N-1$, i.e., all pairs interfere with each other, we can prove that
TDMA achieves better system performance than the GF sequence scheme.

\begin{proposition} \label{prop:D-N-1-TDMA-better-than-GF}
When $D = N-1$ and $N$ is a prime power, TDMA achieves larger or equal average system timely throughput than the GF sequence scheme.
\end{proposition}
\ifx \ISTR \undefined
\else
\begin{IEEEproof}
Please see Appendix~\ref{app:proof-of-lem-D-N-1-TDMA-better-than-GF}.
\end{IEEEproof}
\fi

We remark that although the result in Proposition~\ref{prop:D-N-1-TDMA-better-than-GF} is heuristically expected, its proof is quite involved.

\vspace{-0.2cm}

\subsection{Comparison among TDMA, the GF Sequence Scheme, and the Combination Sequence  Scheme for $D=1$} \label{subsec:compare-seqs-for-D=1}
\ifx \ISTR \undefined
{Due to the page limit, we move the detailed comparison for different schemes when $D=1$ to our technical report \cite{TR}.}
\else
In the case of $D=1$, the sequence periods of TDMA, the GF sequence scheme and
the combination sequence scheme are $N$, $q^2(1,N)$ and $L^{\min}(1,N)$, respectively. Proposition~\ref{prop:min-L-for-D=1}
shows that $L^{\min}(1,N) \le \min\{N,q^2(1,N)\}$.
Namely, the combination sequence scheme
has the shortest period among them. Therefore, according to  Lemma~\ref{lem:R-1-R-2-decrease-with-L},
the combination sequence scheme has the largest lower bound of average system timely throughput.
To compare $q^2(1,N)$ and $N$, we note that $q(1,N)$
is the smallest prime power satisfying \eqref{equ:min-q-GF} with $D=1$.
We prove the following result.

\begin{theorem} \label{thm:compare-q-1-N-and-N}
$q^2(1,N) > N$ when $N \le 8$;
$q^2(1,N) = N$ when $N = 9$;
and $q^2(1,N) < N$ when $N \ge 10$.
In addition, when $N \ge 10$, the GF sequence scheme achieves larger or equal average system timely throughput than TDMA.
\end{theorem}
\ifx \ISTR \undefined
\else
\begin{IEEEproof}
Please see Appendix~\ref{app:proof-of-thm-compare-q-1-N-and-N}.
\end{IEEEproof}
\fi

Therefore, when $N \ge 10$ (resp. $N \le 8$), the period of TDMA  is larger than (resp. smaller than) the period of the GF sequence scheme.
Again we remark that TDMA only guarantees exactly one collision-free slot for each sequence in each period,
while the GF sequence scheme and the combination sequence scheme could have \emph{at least} one collision-free slot in each period.
Therefore, the sequence period does not directly reflect the actual average system timely throughput.
We will compare their actual performance by simulations in Sec.~\ref{sec:simulation}.
\fi

\vspace{-0.2cm}

\section{Simulations} \label{sec:simulation}

In this section, we compare the performance of different topology-transparent distributed schemes by simulations.
\ifx \ISTR \undefined
{Due to the page limit, we present key simulations here. Please refer to our technical report \cite{TR}
for effects of more system parameters, extensions of our model, and simulations for a practical MANET environment.}
\else

In Sec.~\ref{subsec:sim-compare-the-emp}, we compare our theoretical analysis and the empirical results, as well as evaluating
the effect of topology density $D$.
We then evaluate the effect of channel quality in Sec.~\ref{subsec:sim-effect-of-p} and the effect of frame length $T$ in Sec.~\ref{subsec:sim-effect-of-T}.
Finally, in Sec.~\ref{subsec:sim-compare-schemes-for-D=1}, we compare different schemes for the special case of $D=1$.
{
Sec.~\ref{subsec:sim-roubustness} evaluates the robustness of the proposed schemes when the number of users (resp. the topology density) exceeds the predetermined
$N$ (resp. $D$). In addition to frame-synchronized traffic pattern, Sec.~\ref{subsec:sim-poisson} evaluates the performance of the proposed schemes under a poisson-arrival
traffic pattern. Finally, Sec.~\ref{subsec:sim-feedback} shows the performance improvement if the feedback information is available.
}
\fi

\ifx \ISTR \undefined
\else
\subsection{Theoretical Analysis v.s. Empirical Results} \label{subsec:sim-compare-the-emp}
\fi


\ifx \ISTR \undefined

\begin{figure}[t]
  \centering
  \subfigure[$T=30$]{
    \label{fig:effect-D1} 
    \includegraphics[width=0.48\linewidth]{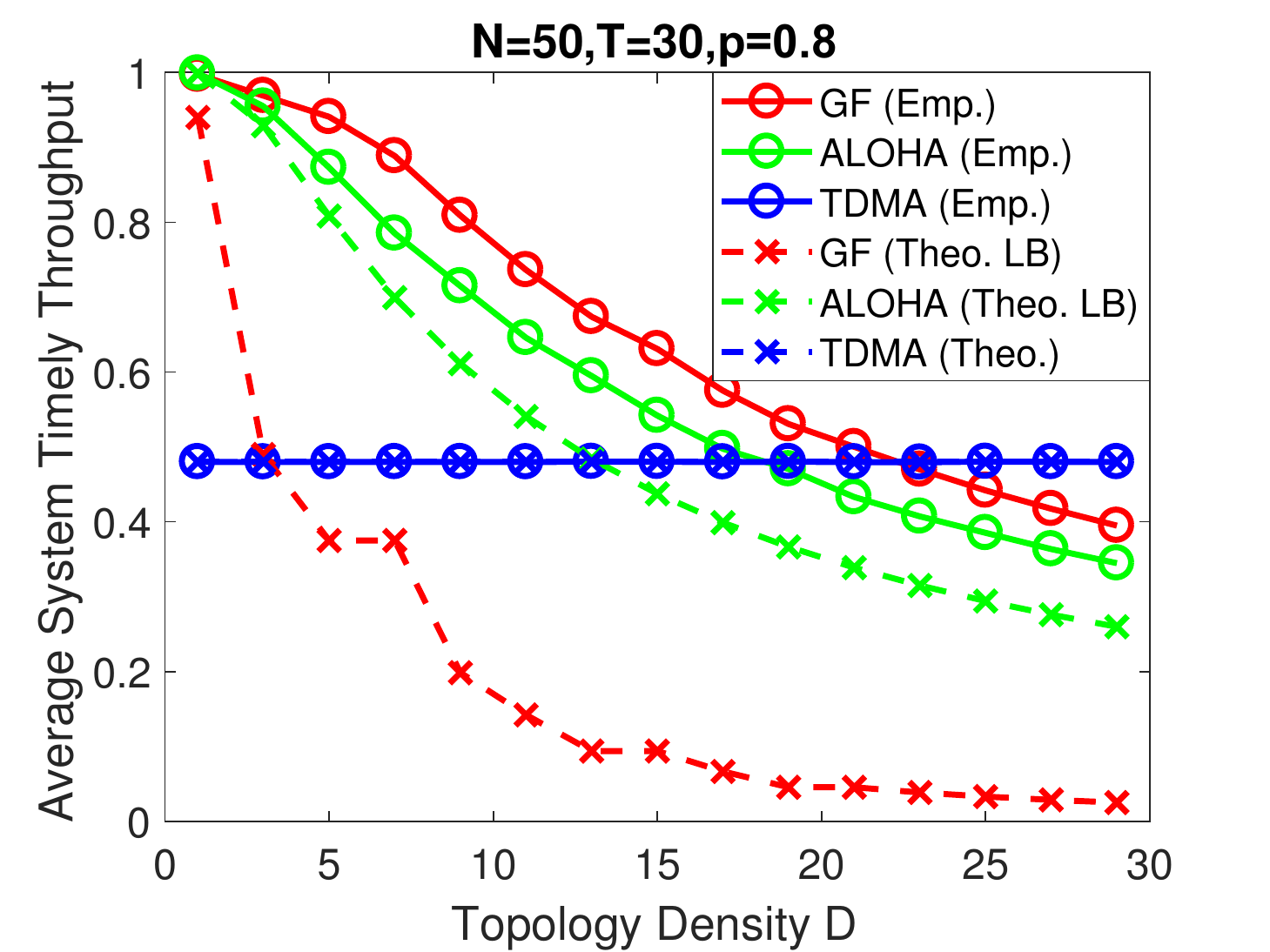}}
    \subfigure[$T=70$]{
    \label{fig:effect-D2} 
    \includegraphics[width=0.48\linewidth]{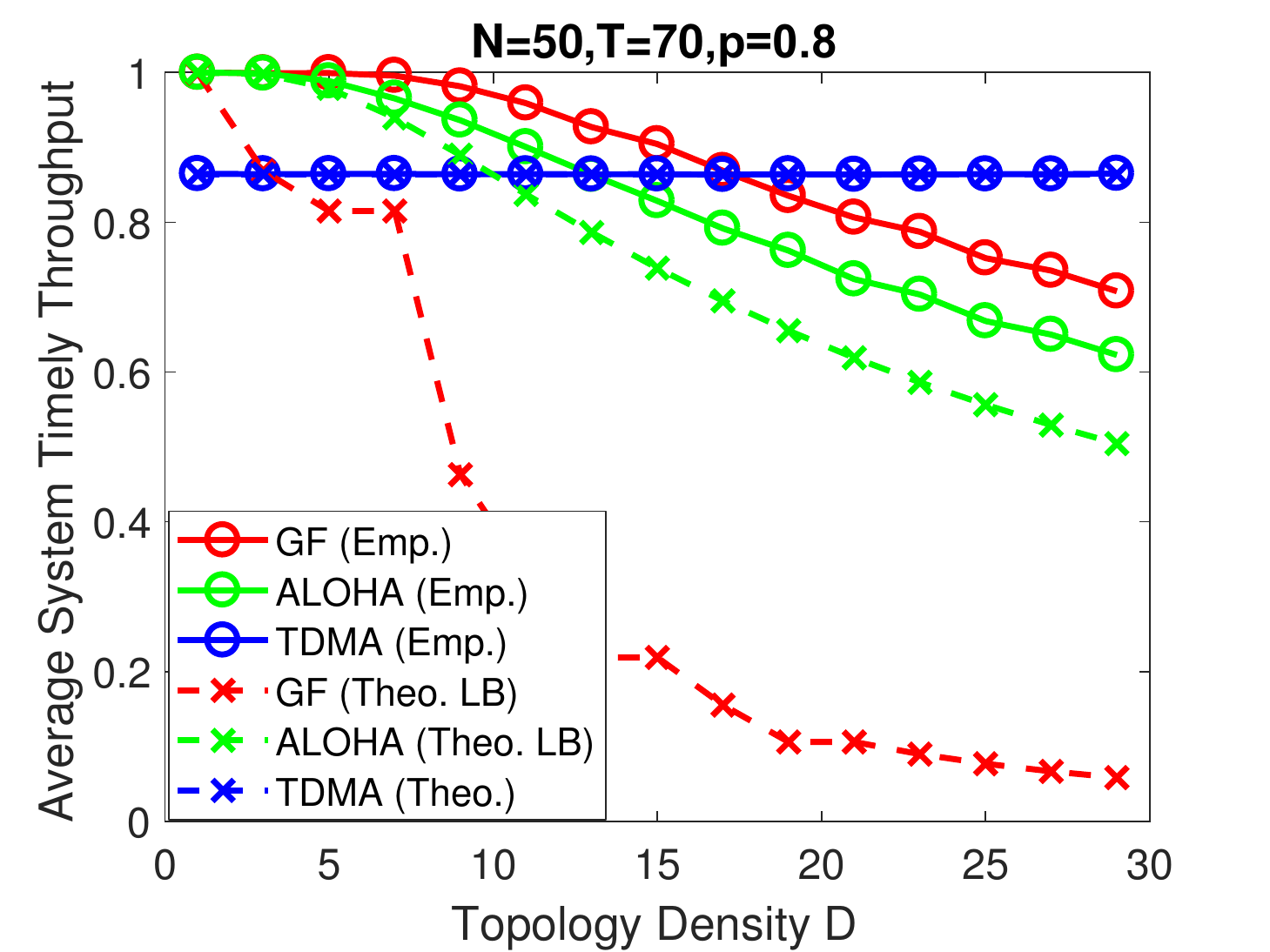}}
  \caption{Compare the empirical  value (Emp.) with theoretical value (Theo.) or theoretical lower bound (Theo. LB), and evaluate the effect of topology density $D$. \label{fig:effect-D} }
\end{figure}
\else
\begin{figure}[t]
  \centering
  \subfigure[$T=30$]{
    \label{fig:effect-D1} 
    \includegraphics[width=0.8\linewidth]{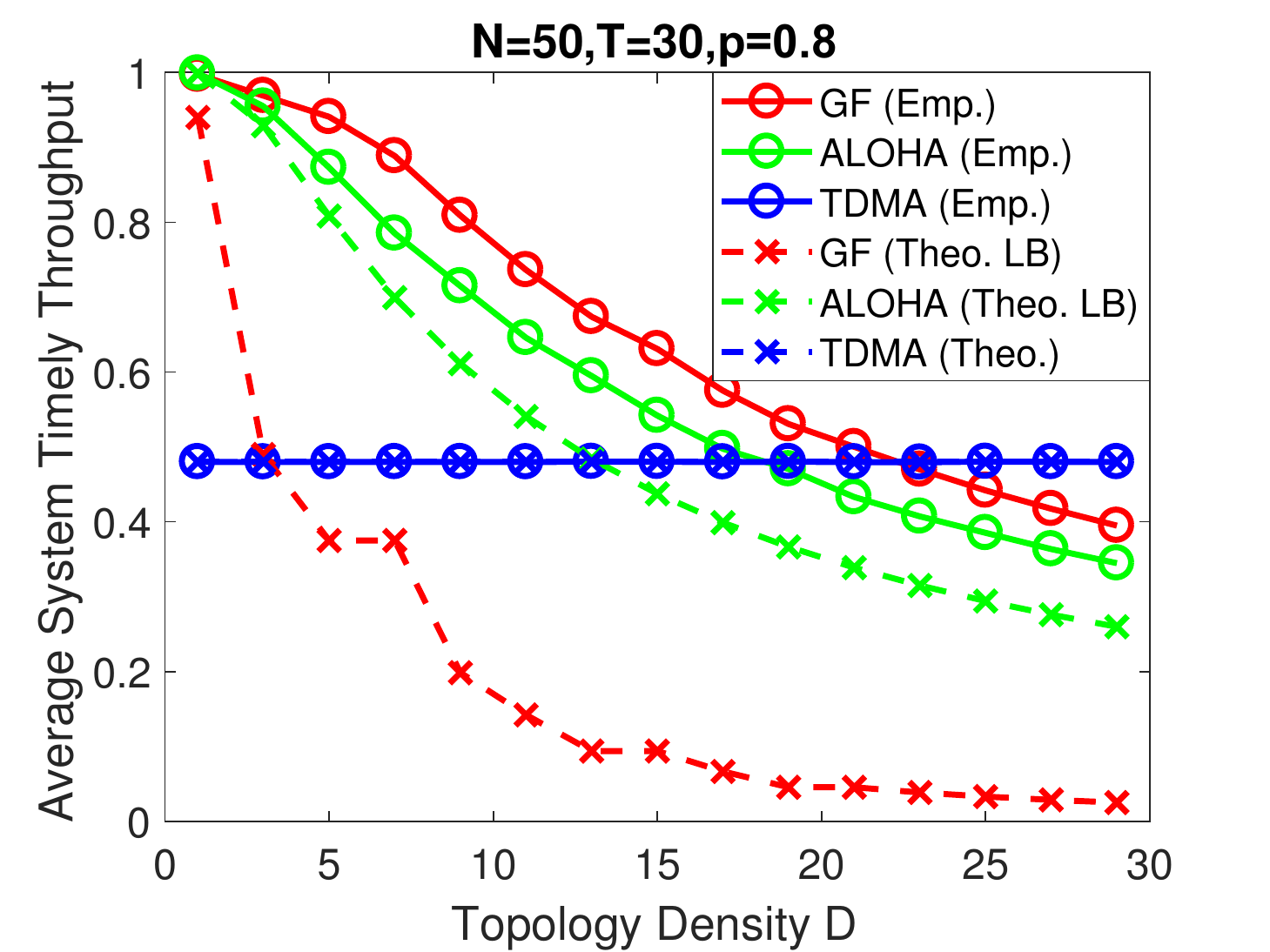}}
    \\ 
    \subfigure[$T=70$]{
    \label{fig:effect-D2} 
    \includegraphics[width=0.8\linewidth]{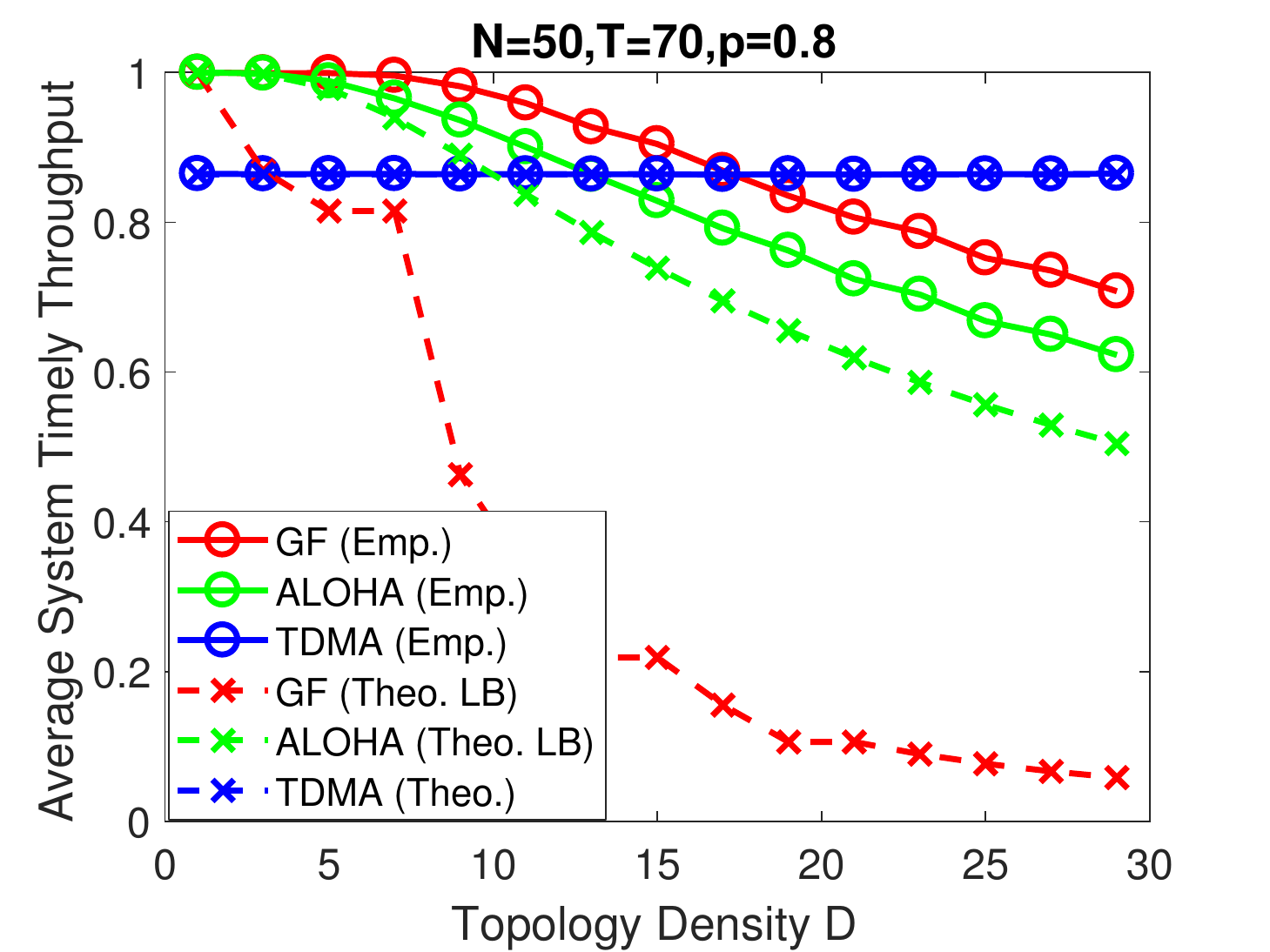}}
  \caption{Compare the empirical  value (Emp.) with theoretical value (Theo.) or theoretical lower bound (Theo. LB), and evaluate the effect of topology density $D$. \label{fig:effect-D} }
\end{figure}

\fi

In terms of average system  timely throughput, we compare theoretical value (or theoretical lower bound)
and the empirical value.
{
We consider $N=50$, $T=30$ or $70$ and $p_i=p=0.8, \forall i\in \{1,2,\cdots,N\}$.
For each topology density $D \in \{1,3,5,\cdots,29 \}$, we randomly generate 100 different network topologies and then calculate the mean value of average system timely throughput for all the 100 topologies.}

\ifx \ISTR \undefined
Fig.~\ref{fig:effect-D} shows theoretical lower bound
\else
Fig.~\ref{fig:effect-D} shows theoretical lower bound in \eqref{equ:avg-system-throughput-ALOHA}
\fi
and empirical result of ALOHA, theoretical result  in \eqref{equ:avg-R-TDMA-1} and \eqref{equ:avg-R-TDMA-2}
and the empirical result of TDMA, and theoretical lower bound  in \eqref{equ:avg-R-GF-1} and \eqref{equ:avg-R-GF-2}
and the empirical result of the GF sequence scheme.

From Fig.~\ref{fig:effect-D}, we can observe that the empirical  performance of TDMA matches well with theoretical result, confirming
the correctness of \eqref{equ:avg-R-TDMA-1} (when $N=50 > T=30$ as shown in Fig.~\ref{fig:effect-D1})
and \eqref{equ:avg-R-TDMA-2} (when $N=50 < T=70$ as shown in Fig.~\ref{fig:effect-D2}). In addition, we can see that
the empirical average system timely throughput of ALOHA is a little bit better than the corresponding theoretical lower bound.
The reason is that the number of interferers for any receiver in the average sense is smaller than $D$, while we derive theoretical lower bound in
\ifx \ISTR \undefined
\else
\eqref{equ:avg-system-throughput-ALOHA}
\fi
based on the maximum number of interferers, i.e., $D$.
The empirical average system timely throughput of the GF sequence scheme is much better than the corresponding theoretical lower bound.
The reason is that the number of collision-free slots in any period of the GF sequence scheme could be much larger than one
as its code weight (number of 1 in a period) is $q>1$,
while we derive theoretical lower bound in \eqref{equ:avg-R-GF-1} and \eqref{equ:avg-R-GF-2} based on the assumption that
each sequence has only one collision-free slot in any period.

In addition, we can see that the  performance of both ALOHA and the GF sequence scheme degrades when~$D$ increases.
This is because larger~$D$ implies more interferers for a receiver and thus ALOHA is more  vulnerable to collision and the GF sequence scheme requires longer sequence period.
The performance of TDMA does not change with respect to~$D$, confirming  our remark in the last paragraph of Sec.~\ref{subsec:TDMA}.
We further note that the  performance of the GF sequence scheme and ALOHA is better than TDMA when~$D$ is small. This is because
both the GF sequence scheme and ALOHA can exploit the sparsity of the network while TDMA cannot. But when~$D$ is large, TDMA dominates others because
the  performance of ALOHA and the GF sequence scheme degrades when~$D$ increases while that of TDMA does not change.
This confirms our analysis in Sec.~\ref{subsec:compare-aloha-TDMA} and Sec.~\ref{subsec:compare-GF-and-TDMA}.


\ifx \ISTR \undefined
\else
\subsection{Effect of Channel Quality $p_i$} \label{subsec:sim-effect-of-p}
\fi

\ifx \ISTR \undefined
\begin{figure}[t]
  \centering
  \subfigure[$D=3$]{
    \label{fig:effect-p1} 
    \includegraphics[width=0.48\linewidth]{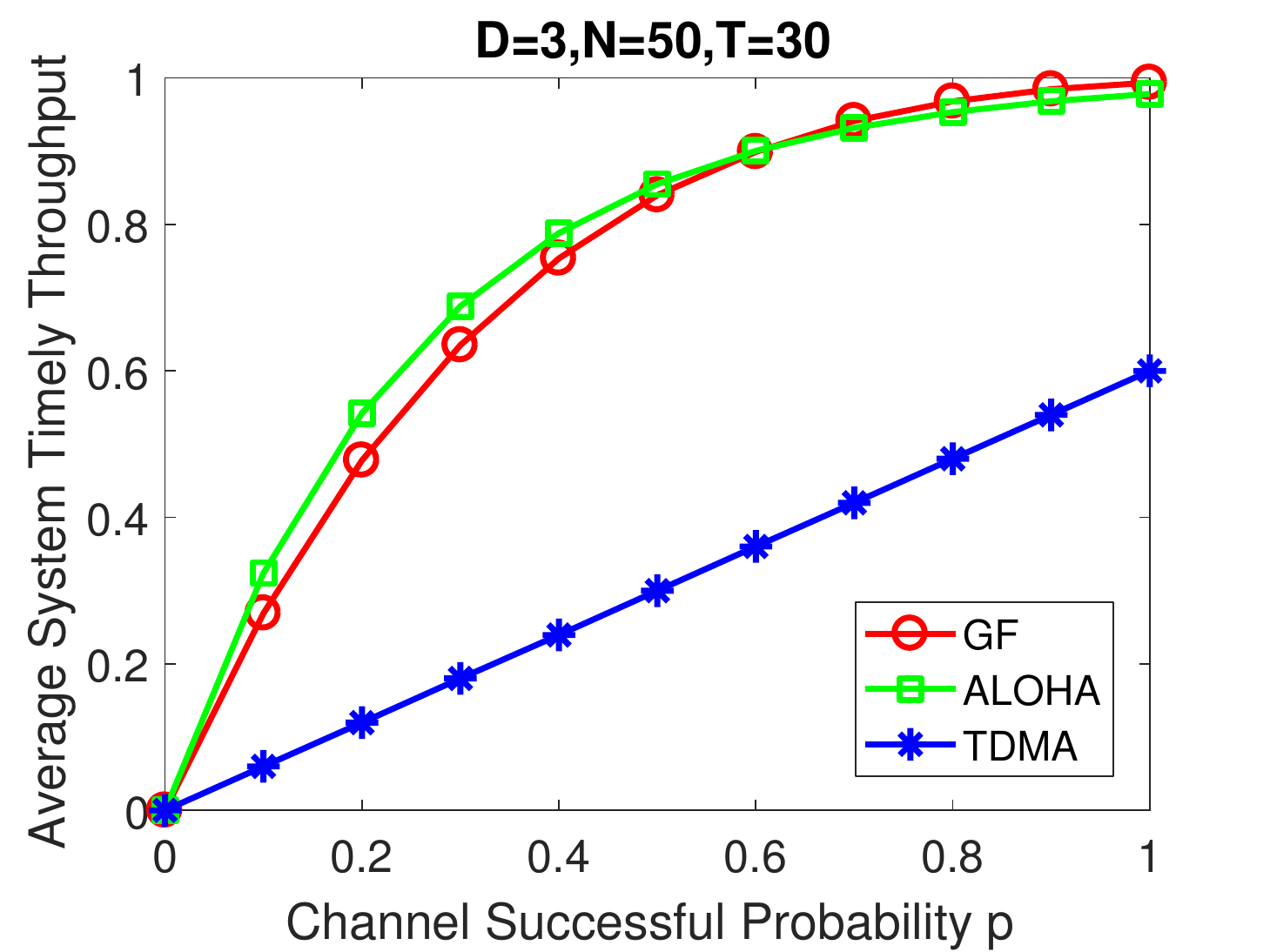}}
    \subfigure[$D=30$]{
    \label{fig:effect-p2} 
    \includegraphics[width=0.48\linewidth]{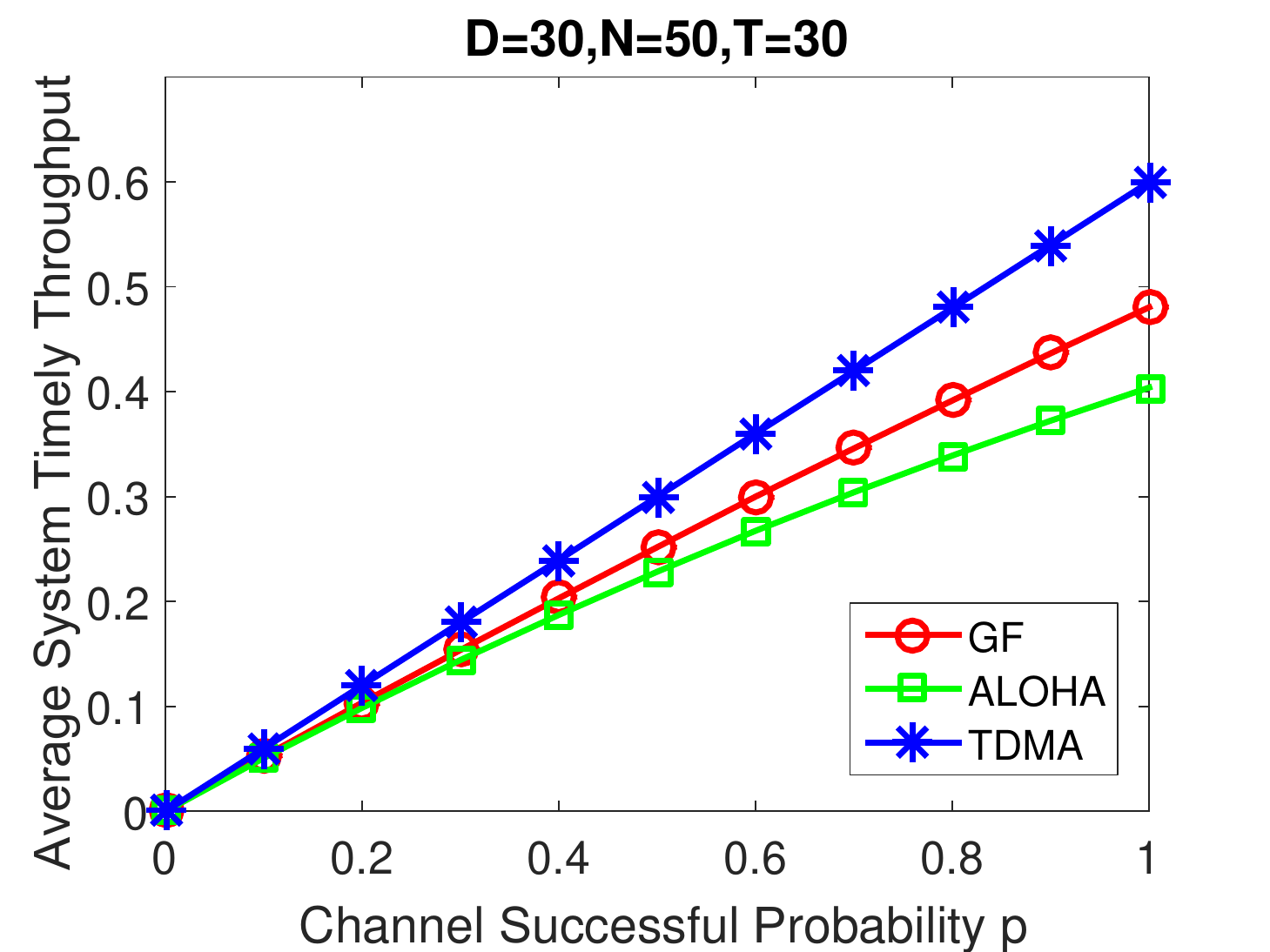}}
  \caption{Effect of channel quality. \label{fig:effect-p} } 
\end{figure}
\else
\begin{figure}[t]
  \centering
  \subfigure[$D=3$]{
    \label{fig:effect-p1} 
    \includegraphics[width=0.8\linewidth]{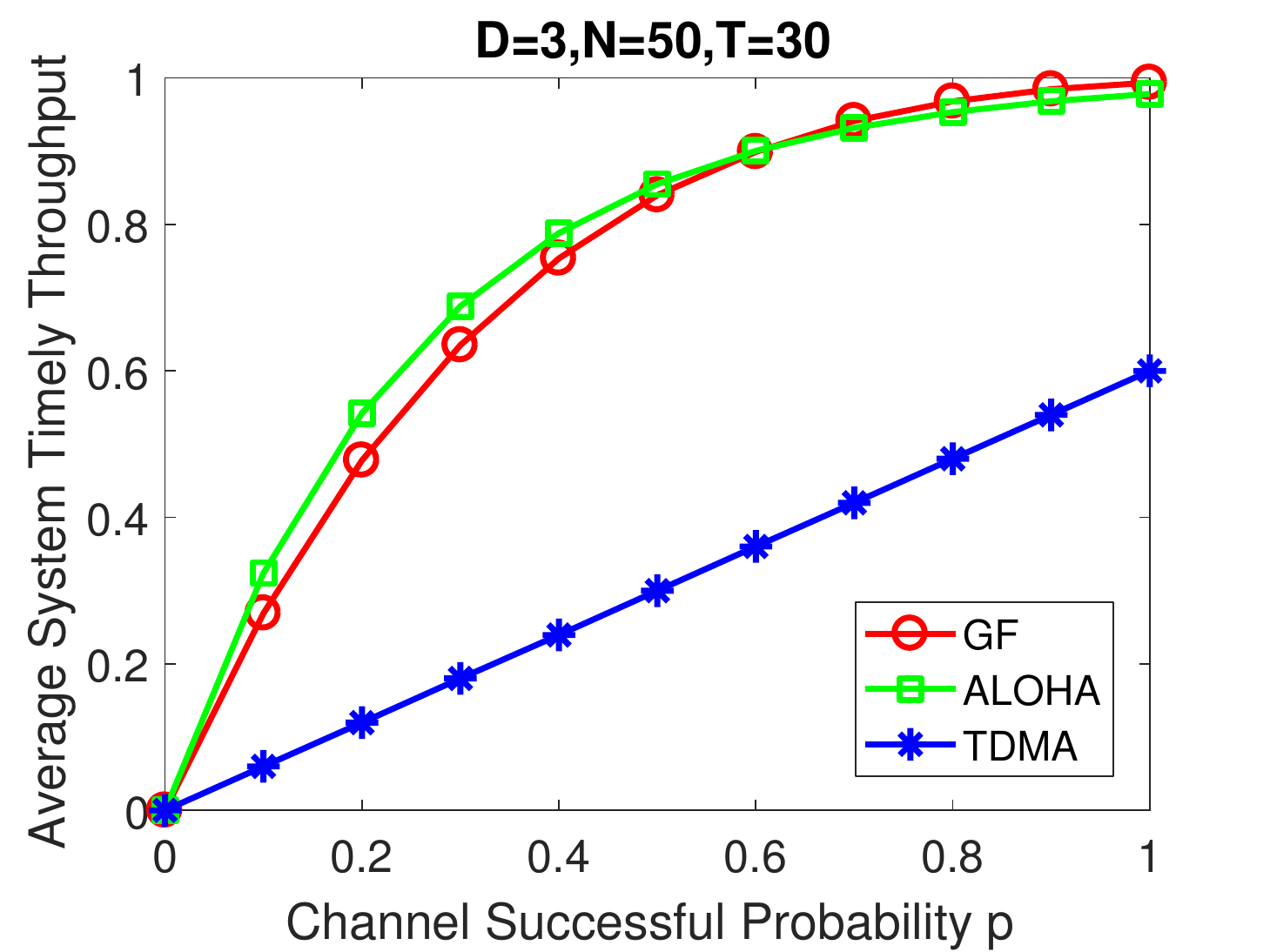}}
    \\ 
    \subfigure[$D=30$]{
    \label{fig:effect-p2} 
    \includegraphics[width=0.8\linewidth]{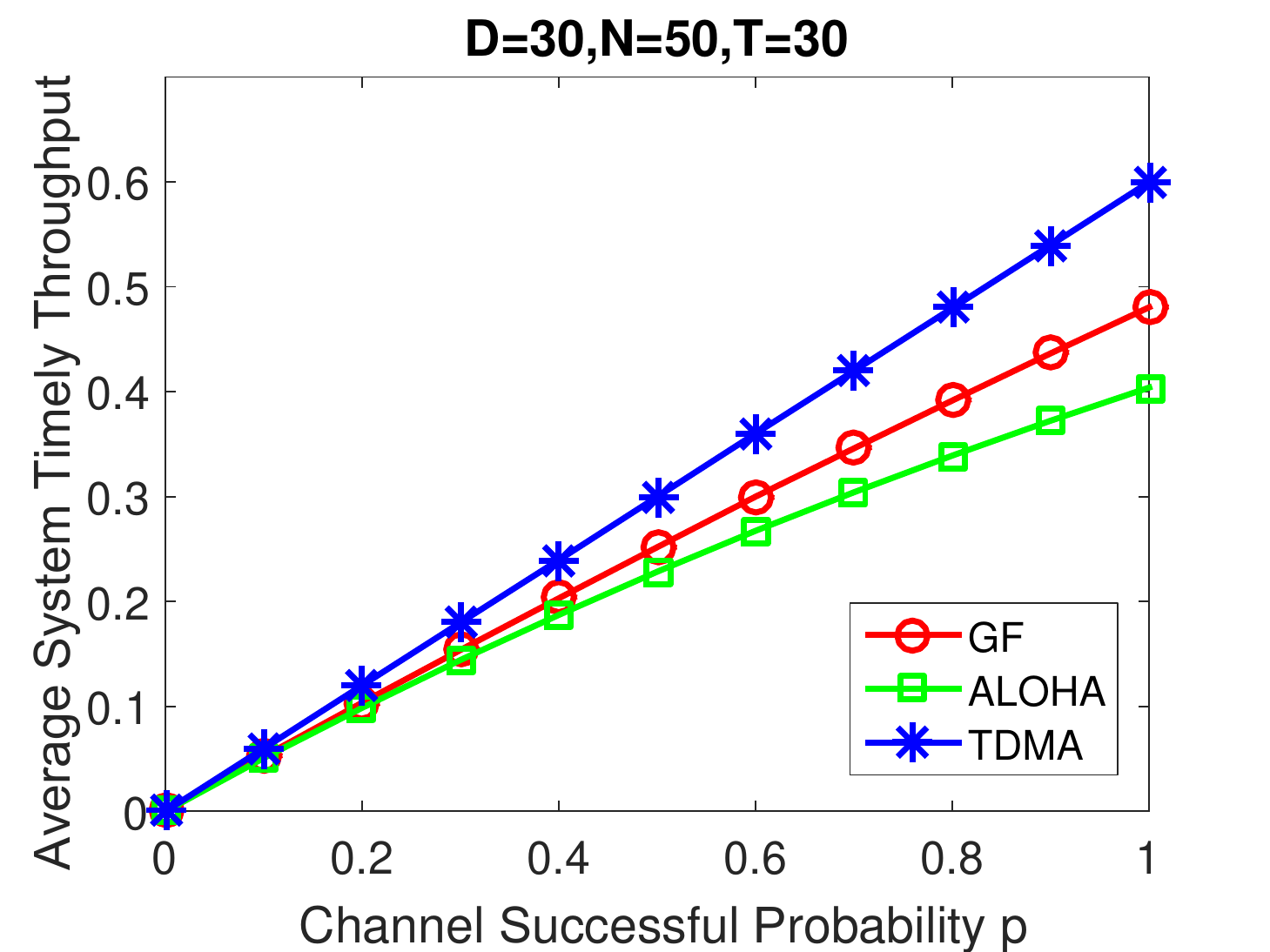}}
  \caption{Effect of channel quality. \label{fig:effect-p} } 
\end{figure}
\fi

We also evaluate the effect of channel quality $ p_i $.
{
We assume that all pairs have the same channel quality, i.e., $p_i=p, \forall i\in \{1,2,\cdots,N\}$,
where $p$ varies from 0 to 1. We set $N=50$, $T=30$, and $D=3$ or 30.
For each $D$, we randomly generate 100 topologies.
For each $p$,  we run 100 times for each of the 100 topologies, and then calculate the mean value of average system timely throughput.
}The  throughput performance of ALOHA, TDMA and the GF sequence scheme is shown in Fig.~\ref{fig:effect-p}.
We can see that in all schemes, better channel quality leads to larger average system timely throughput. This is an obvious result.
Again, similar to the analysis for Fig.~\ref{fig:effect-D}, the  performance of the GF sequence scheme
and ALOHA is better than TDMA when $D$ is small ($D=3$), while TDMA has the best  performance when $D$ is large ($D=30$).

In addition, when $D$ is small  and channel quality $p$ is small,
ALOHA is better than the GF sequence scheme. We have also carried out many other instances to confirm this observation.
This shows that although both ALOHA and the GF sequence scheme improve their performance when $D$ decreases, the improvement of ALOHA outperforms that of the GF sequence scheme when the channel quality is low.

\ifx \ISTR \undefined
\else

\begin{figure}[t]
  \centering
  \subfigure[$D=1$]{
    \label{fig:effect-T-1} 
    \includegraphics[width=0.8\linewidth]{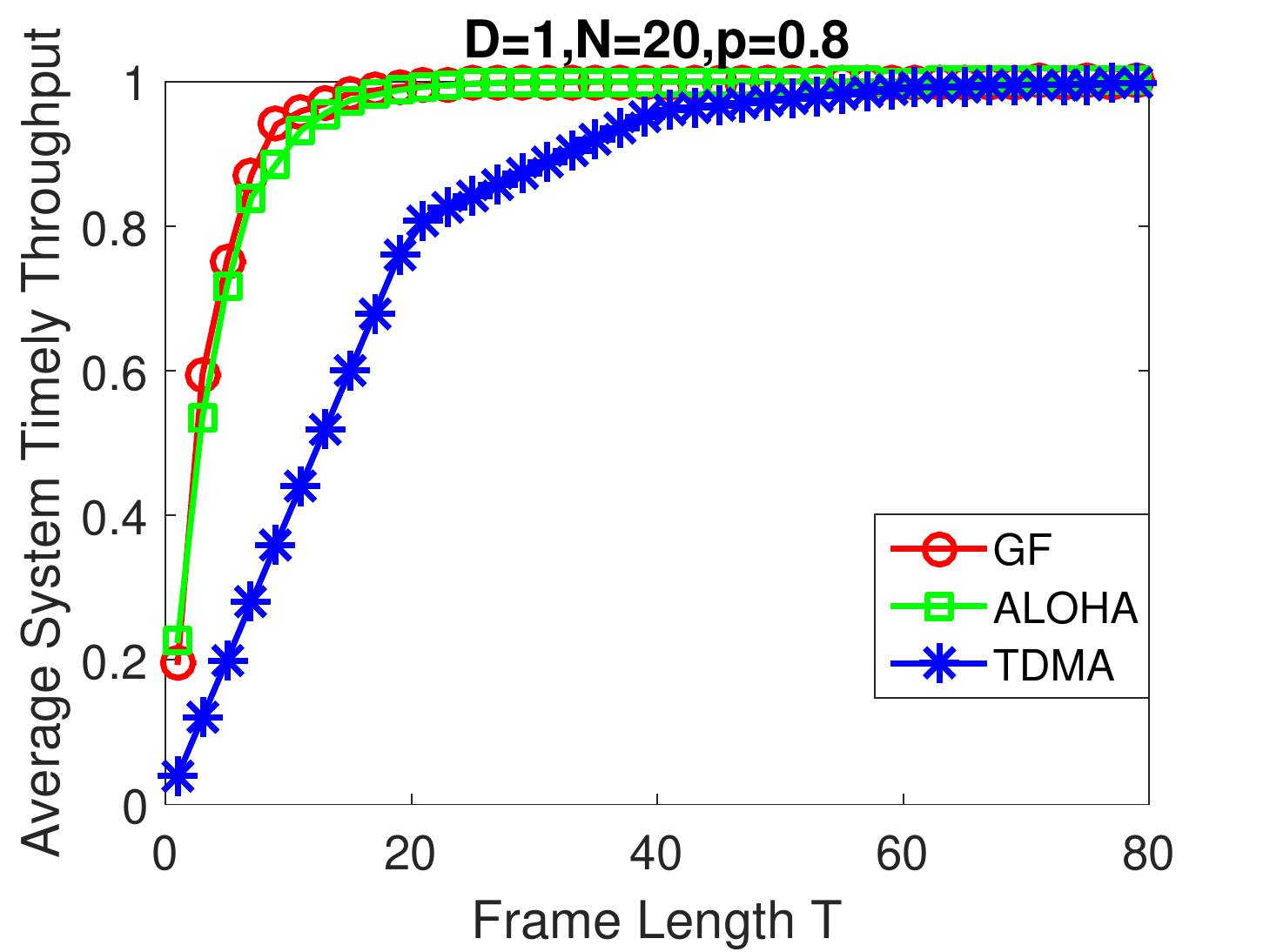}}
    \\ 
    \subfigure[$D=10$ ]{
    \label{fig:effect-T-2} 
    \includegraphics[width=0.8\linewidth]{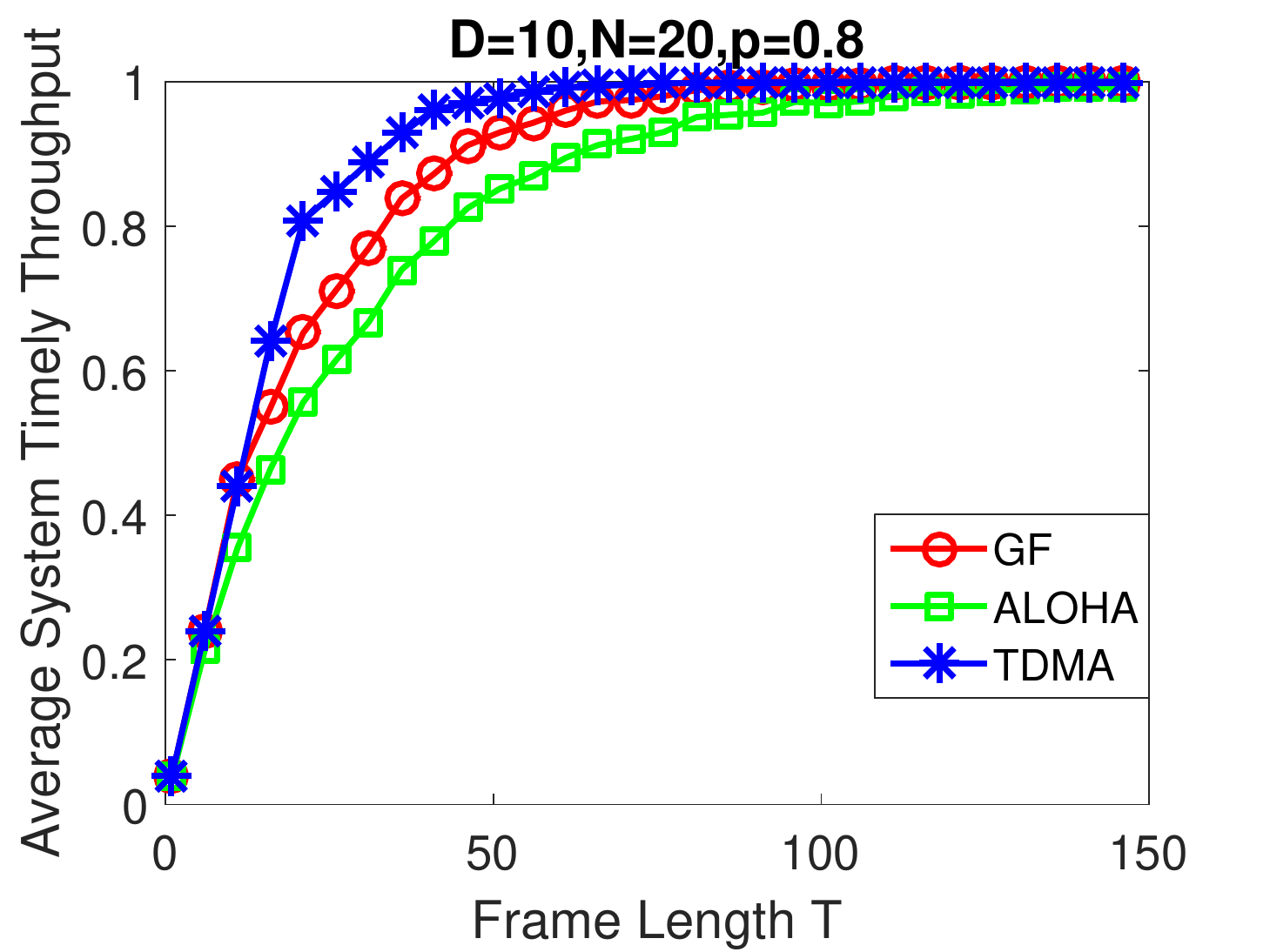}}
  \caption{Effect of frame length. \label{fig:effect-T}}\label{fig:effect-T}
\end{figure}

\begin{figure}[t]
  \centering
  \subfigure[$T=10$]{
    \label{fig:throughput-comparision-D=1-1} 
    \includegraphics[width=0.8\linewidth]{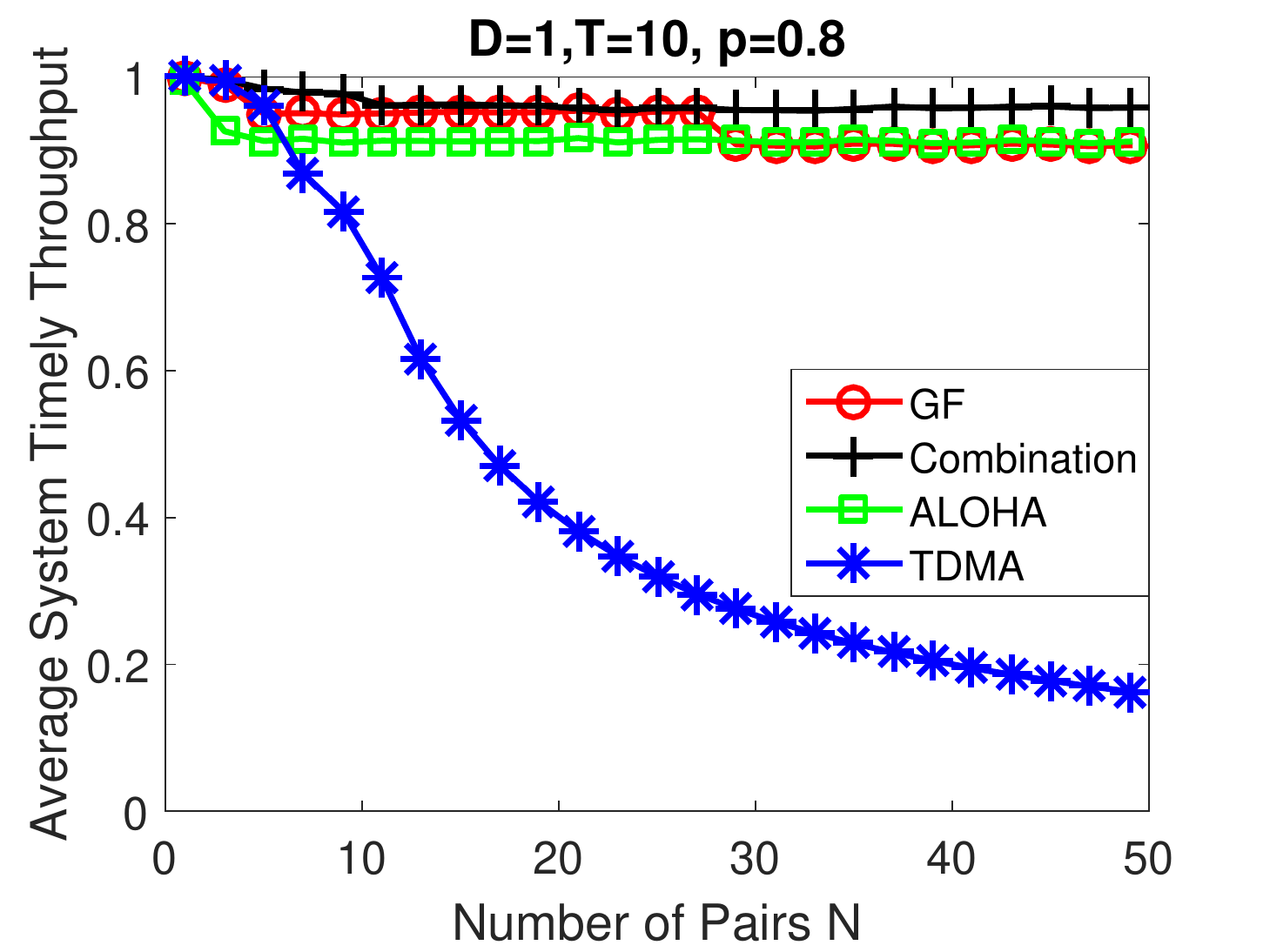}}
    \\ 
    \subfigure[$N=10$]{
    \label{fig:throughput-comparision-D=1-2} 
    \includegraphics[width=0.8\linewidth]{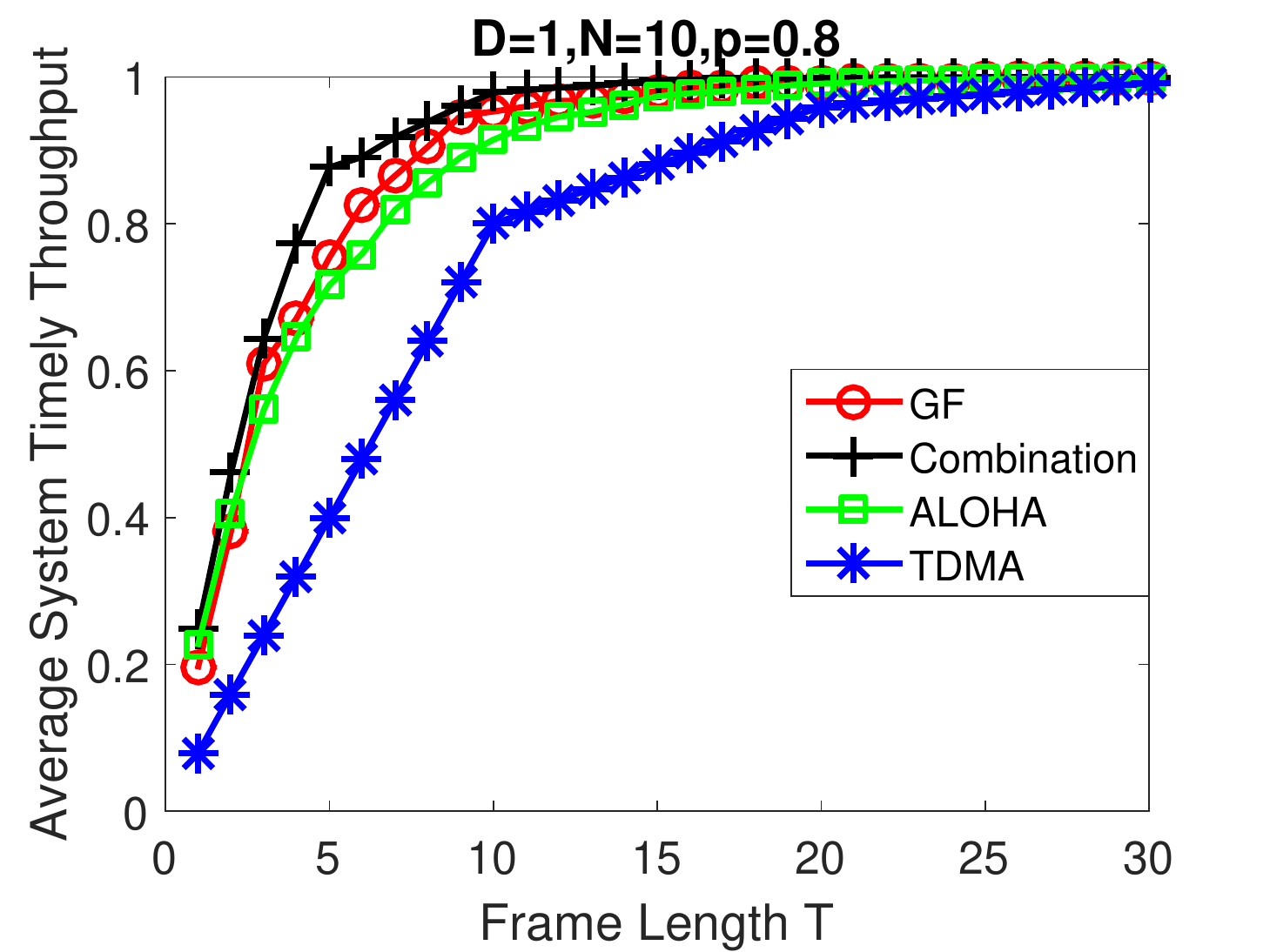}}
  \caption{Compare different schemes for the special case of $D=1$. \label{fig:throughput-comparision-D=1} }
\end{figure}

\subsection{Effect of Hard Deadline/Frame Length $T$} \label{subsec:sim-effect-of-T}


In this part, we evaluate the effect of hard deadline/frame length $T$.
{
We set $N=20$, $D=1$ or $10$, and $p_i=p=0.8, \forall i\in \{1,2,\cdots,N \}$. For each $D$, 100 topologies are randomly generated.
For each $T$, we calculate the mean value of average system timely throughput  for each of the 100 topologies in 100 runs.
}
The result is shown in Fig.~\ref{fig:effect-T}. We can see that the average system timely throughput of
all three schemes (ALOHA, TDMA and the GF sequence scheme) increases as $T$ increases. This is because larger frame length can allow
transmitters to have more slots to transmit packets in a frame.

In the case of $D=1$, the period length of the GF sequence scheme is equal to 9, which is much shorter than that of TDMA. From Fig.~\ref{fig:effect-T-1}, we can see that the GF sequence scheme is faster than TDMA to converge to the maximum average system timely throughput as $T$ increases.
On the other hand, in the case of $D=10$, the period length of the GF sequence scheme is equal to 121, which is much longer than that of TDMA. From Fig.~\ref{fig:effect-T-2}, we can observe that TDMA is faster than the GF sequence scheme to converge to the maximum average system timely throughput.


\subsection{Comparison among Different Schemes when $D=1$} \label{subsec:sim-compare-schemes-for-D=1}
We compare the four  schemes (ALOHA, TDMA, the GF sequence scheme and the combination sequence scheme)
for the special case of $D=1$.
{
Two cases with $p_i=p=0.8, \forall i\in \{1,2,\cdots,N \}$, are investigated. In the first case, we fix frame length $ T=10 $ and change the number of pairs $ N $.  In the other one, we fix $N=10$ and change $T$. For each $N$ in the first case and each $T$ in the second case, we calculate the mean value of average system timely throughput for 100 randomly generated topologies each with 100 runs.
}
The result is shown in Fig.~\ref{fig:throughput-comparision-D=1}. As we can see,
no matter whether  we fix  $T$ but vary  $N$ as shown in Fig.~\ref{fig:throughput-comparision-D=1-1} or we fix  $N$ but vary  $T$ as shown in Fig.~\ref{fig:throughput-comparision-D=1-2},
the combination sequence scheme has the best performance, which is in line with our analysis in Sec.~\ref{subsec:compare-seqs-for-D=1}.
Though we designed the combination sequence scheme for the special case of $D=1$,
we can also apply it for the cases of $D>1$ to see its practical performance.
However, according to our independent simulations (omitted here due to the space limitation),
we find that the combination sequence scheme is worse than the GF sequence scheme and ALOHA when $D>1$. Thus,
the combination sequence scheme should only be used for the special case of $D=1$.

{
\subsection{Robustness of Our Schemes} \label{subsec:sim-roubustness}
In this paper, similar to existing literatures on topology-transparent scheduling \cite{chlamtac1994making,su2015topology,kar2017survey}, we assume that the number
of pairs, i.e., $N$, is fixed and then design sequence schemes based on this given $N$. However, in practice, it is possible that new pairs enter the network and/or
existing pairs leave the network. It is necessary to show the robustness of our schemes under this situation.
Toward that end, we need to design a practical solution to assign sequences to new pairs\footnote{{The behaviour of new pairs does not change under the ALOHA scheme}.}.
Generally the sequence space of the designed scheme is larger than the number of pairs, i.e., $N$. For example, in GF sequence scheme, the sequence space is of size $q^{k+1} \ge N$ (see Equ. (13)),
which usually holds as an inequality.
Thus, we have extra sequences to be allocated to a new transmitter. On the other hand, if we have exhausted the sequence space when a new pair enters the network,
we can still randomly allocate a sequence from the sequence space to this new transmitter. Although two transmitters have the same sequence now,
the chance that they are close and thus interfere with each other is low. This kind of sequence reuse is similar to the idea of spectrum reuse in cellular networks.
Even though they have chance to interfere with each other, this simple solution is also practical by sacrificing a little bit system performance.
In this part, we apply this practical approach when new pairs enter the network.

In addition, similar to many existing literatures on topology-transparent scheduling design (see \cite{chlamtac1994making,su2015topology,kar2017survey} and the references therein),
we make the assumption that $D$ is known to all nodes and the network density would not exceed $D$ all the time.
Under this assumption, we can design schemes based on $D$, and we reveal insights about which scheme performs better under different conditions.
In practice, the value of $D$ can be estimated from observations for different kinds of network scenarios.
However, due to the mobility of VANET, the network density could exceed the predetermined value $D$. It is also important to evaluate the robustness of our schemes under this situation.

{
To evaluate the robustness of our schemes when  $N$ and $D$ violate the predetermined values, we perform  simulations as follows.
We initially set up a network with $N=50$, $D=10$, $T=50$, and $p_i=p=0.8 \; (\forall i)$.
Then we increase the total number of pairs $N$ from 50 (the predetermined value) to 75 and keep the network density $D$ unchanged.
For each $N$, we generate 100 topologies randomly and run 100 times for each topology. The mean values of average system timely throughput for different $N$'s are plotted in Fig.~\ref{fig:robustness-N}.
We can observe that the throughput performance under both GF and TDMA degrades as $N$ increases,
while the throughput performance under ALOHA largely remains unchanged since it mainly depends on $D$
(as shown in Sec.~\ref{sec:aloha}).
To describe the degradation more accurately, we define changing ratio for a sequence $\{x(t):t=1,2,\cdots\}$ as $\frac{\max\{x(t)\}-\min\{x(t)\}}{x(1)}$. We  calculate that the changing ratio of $N$ is 50\%, and the changing ratios of the system performance of GF and TDMA schemes are  3.2\% and 7.3\%, respectively.
The reason why the throughput reduction under GF is less than that under TDMA is that the GF sequence space size in the case of $ N=50 $ and $D=10$ is 121, which is much larger than the TDMA sequence space size, i.e., 50. The larger the space size, the less the chance that two interfering pairs  choose the same sequence.

In Fig.~\ref{fig:robustness-D}, we increase the network density $D$ from 10 (the predetermined value) to 15 and fix $N=50$.
For each $D$, we generate 100 topologies randomly and run 100 times for each topology, and then calculate the mean value of average system timely throughput. As shown in Fig.~\ref{fig:robustness-D},
when $D$ increases, the throughput performance under both GF and ALOHA degrades since more collisions would occur. The throughput performance under TDMA keeps the same since $D$ has no impact on TDMA (as shown in Sec.~\ref{subsec:TDMA}).
For Fig.~\ref{fig:robustness-D}, we calculate that the changing ratio of $D$ is 50\%, and the changing ratios of the throughput performance of GF and ALOHA schemes are  12.8\%, and 11.8\%, respectively. Compared with the changing ratios caused by $N$, we can observe that the throughput performance is more sensitive to the change of $D$ and is more robust to the change of $N$.

We also simulate the case where both $N$ and $D$ could increase or decrease.
We change the network topology every 50 frames (which we call a big frame).
At the beginning of each big frame, $N$ randomly takes values from 50 to 75, and $D$ randomly takes values from 10 to 15.
The running system average timely throughput and the dynamics of $N$ and $D$ are shown in Fig.~\ref{fig:robustness-D-N}.
In Fig.~\ref{fig:robustness-D-N}, the changing ratios of $N$ and $D$ are 50\% and 50\%, respectively.
However, the changing ratios of the system performance of GF, ALOHA and TDMA schemes are 12\%, 13.4\%, and 11.1\%, respectively.
Thus, when $N$ and $D$ change, the performance of our schemes does not change significantly. That is, our schemes are robust to  $N$ and $D$.
the reason that TDMA scheme shows the smallest variation due to the fact that it is not affected by parameter $D$, whose effect is more significant than the effect of parameter $N$.
}

\begin{figure}
  \centering
  \subfigure[{Robustness of the proposed GF, ALOHA, and TDMA schemes when $N$  increases from the predetermined value 50.}]{
  \label{fig:robustness-N}
  \includegraphics[width=0.8\linewidth]{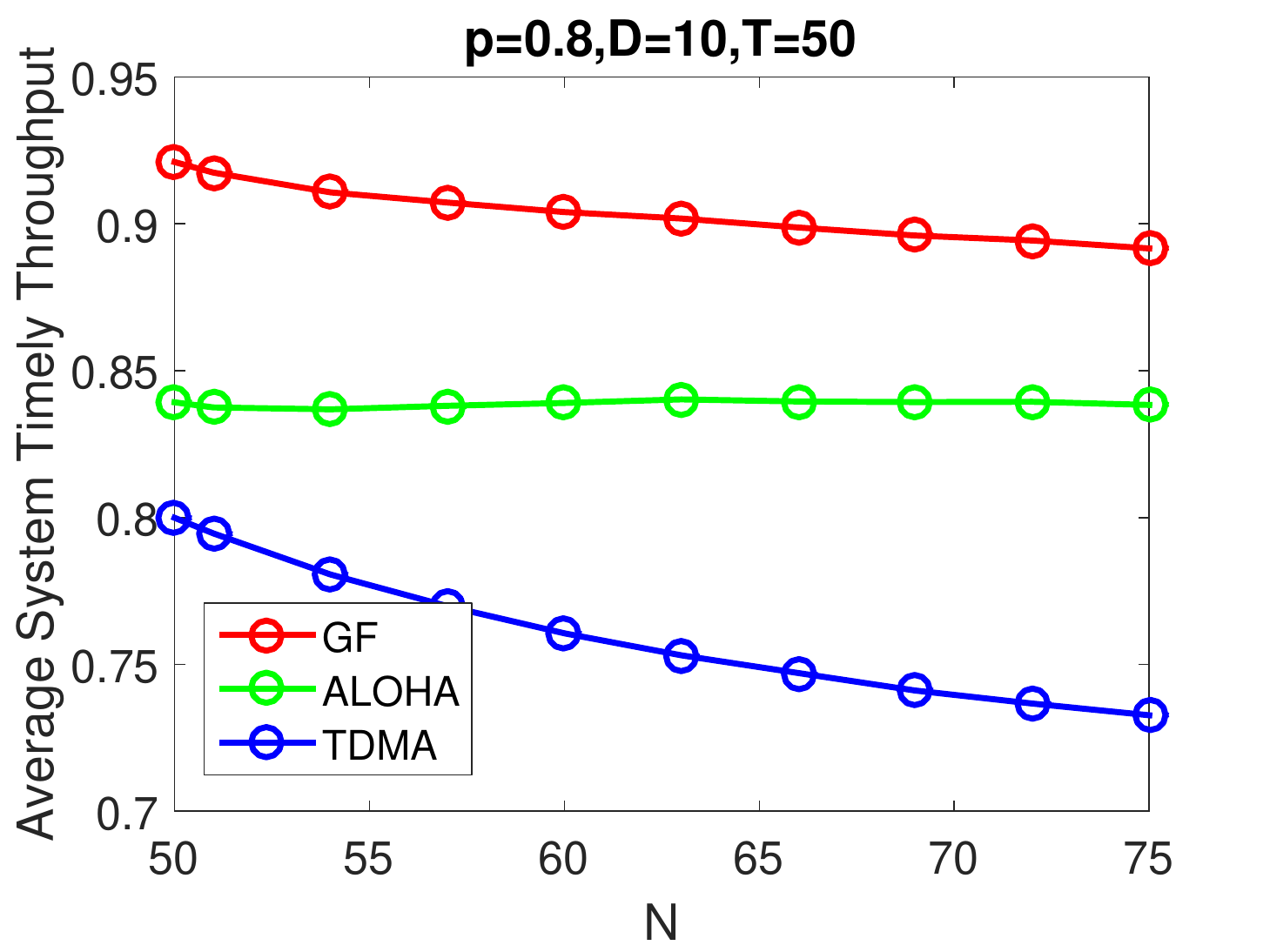}
  }
  \\
  \subfigure[{Robustness of the proposed GF, ALOHA, and TDMA schemes when $D$ increases from the predetermined value 10.}]{
  \label{fig:robustness-D}
  \includegraphics[width=0.8\linewidth]{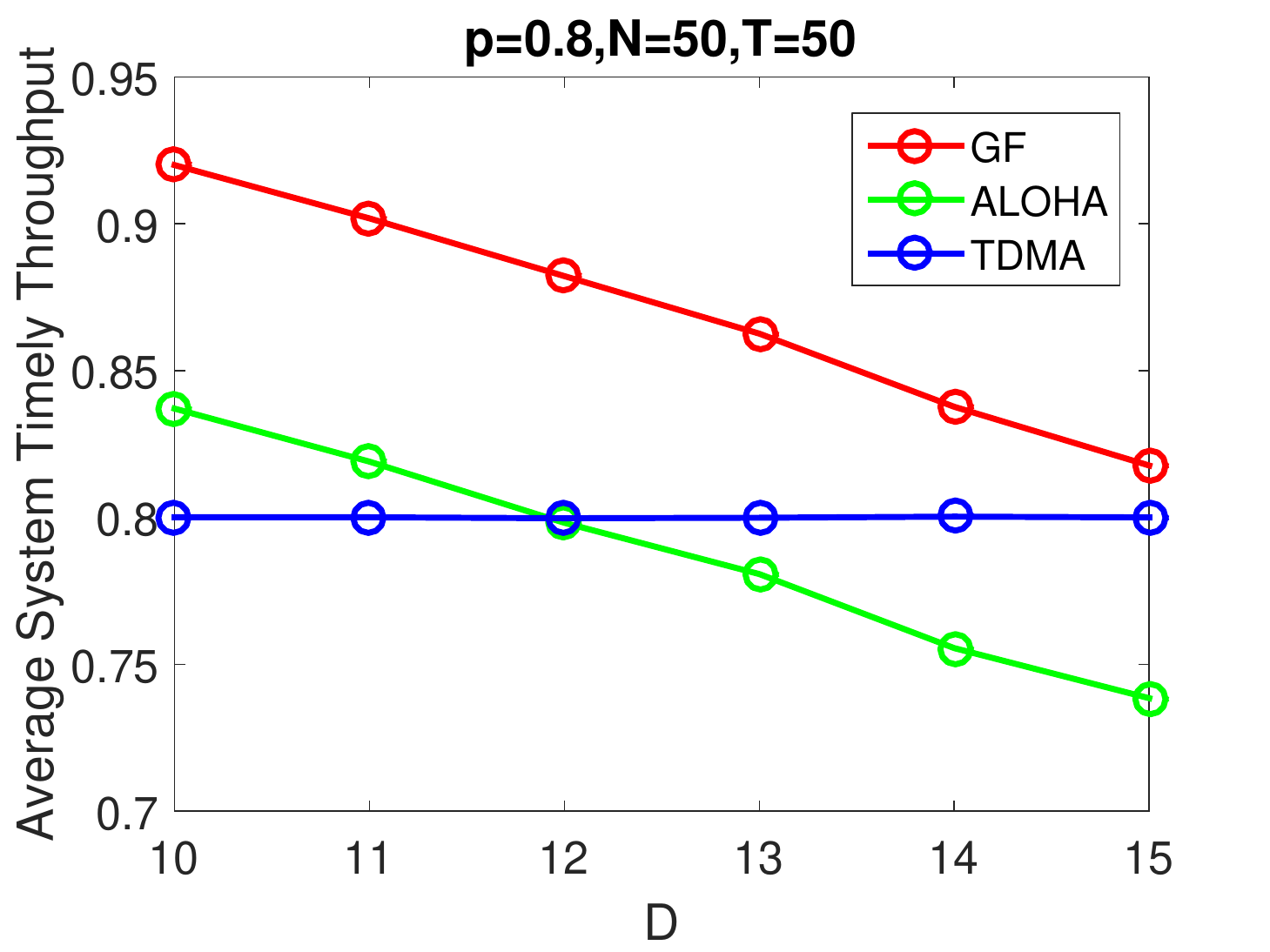}
  }
  \\
  \subfigure[{Robustness of the proposed GF, ALOHA, and TDMA schemes when $N$ and $D$ deviate from their predetermined values.}]{
  \label{fig:robustness-D-N}
  \includegraphics[width=0.8\linewidth]{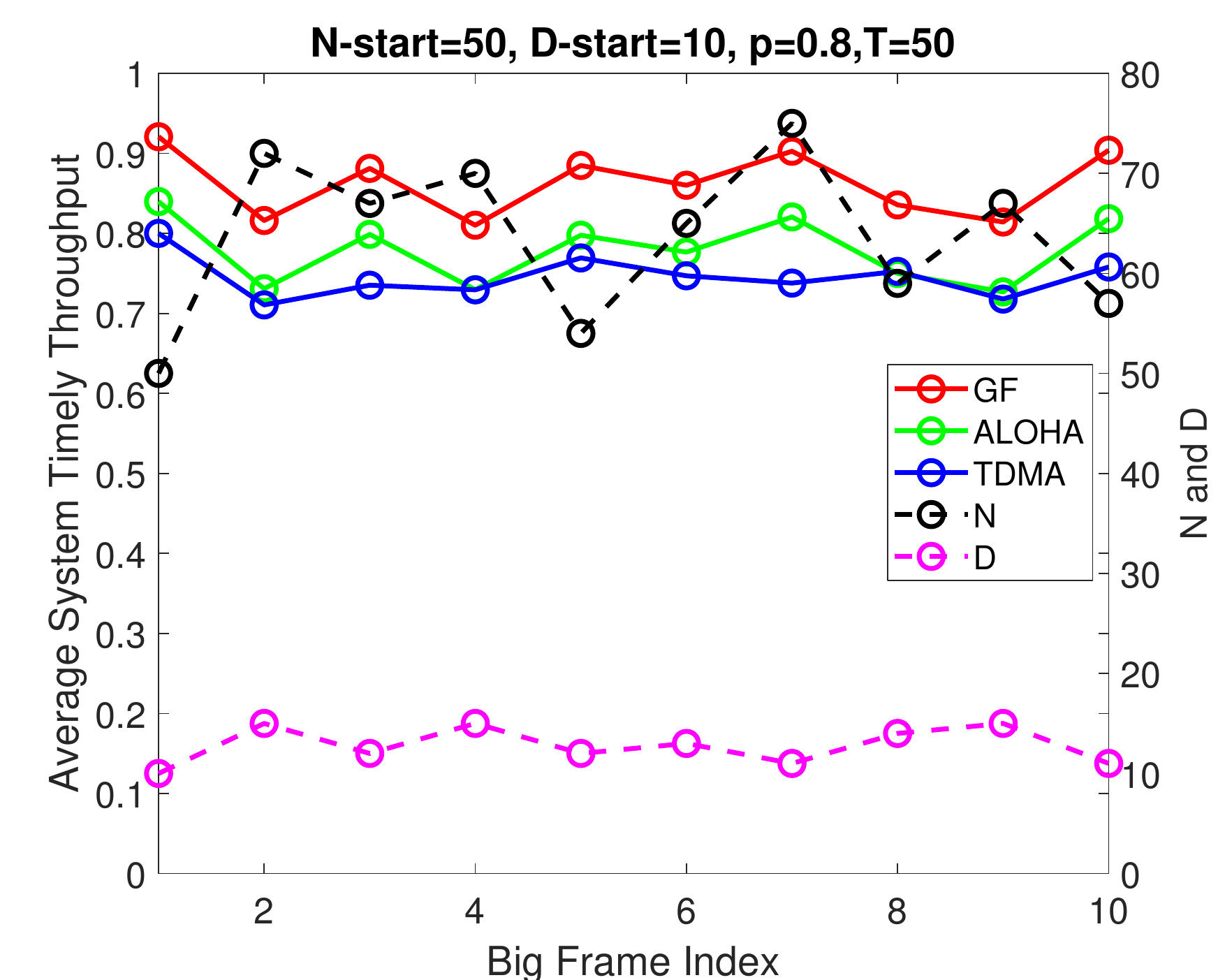}
  }
  \caption{{Robustness of the proposed GF, ALOHA, and TDMA schemes.}}
\end{figure}

\subsection{Poisson Arrival} \label{subsec:sim-poisson}
In this paper, we consider the frame-synchronized traffic pattern, which can find applications in CPSs \cite{kim2012cyber} and NCSs \cite{dengonstability2018}
and is a good starting point to investigate the delay-constrained communications  \cite{hou2009qos,hou2010utility,deng2017timely}.
However, in practice, there are other traffic patterns. For example, poisson arrival is common. It is good to show the performance
of our schemes under this practical traffic pattern. Therefore, we perform a simulation for poisson arrival.
{
We consider a poisson-arrival traffic pattern where the mean of the inter-arrival time is $T=10$. We round the arrival time to an integer in line
with our slotted system. We then apply the proposed GF, ALOHA, and TDMA schemes.
In the setting, $N=20, T=10$, and $D$ is varying.
For each $D$, 100 topologies are randomly generated. Mean values of average system timely throughput for the 100 topologies with $p_i=p=0.8 \; (\forall i)$ and $p_i=p=0.1 \; (\forall i)$ are shown in Fig.~\ref{fig:effect-poisson}.
}
 As we can see, the performance of GF and ALOHA degrades as $D$ increases while
that of TDMA does not change. In addition, GF and ALOHA outperform TDMA when $D$ is small, while TDMA outperforms GF and ALOHA when $D$ is
large. Further, ALOHA is better than GF when $D$ is small and the successful probability $p$ is small. All such observations are the same as
those for frame-synchronized traffic pattern in Sec.~\ref{subsec:sim-compare-the-emp} and Sec.~\ref{subsec:sim-effect-of-p}. All of them conform
our conclusion later in Tab.~\ref{tab:summary}.

\begin{figure}
  \centering
  \includegraphics[width=0.8\linewidth]{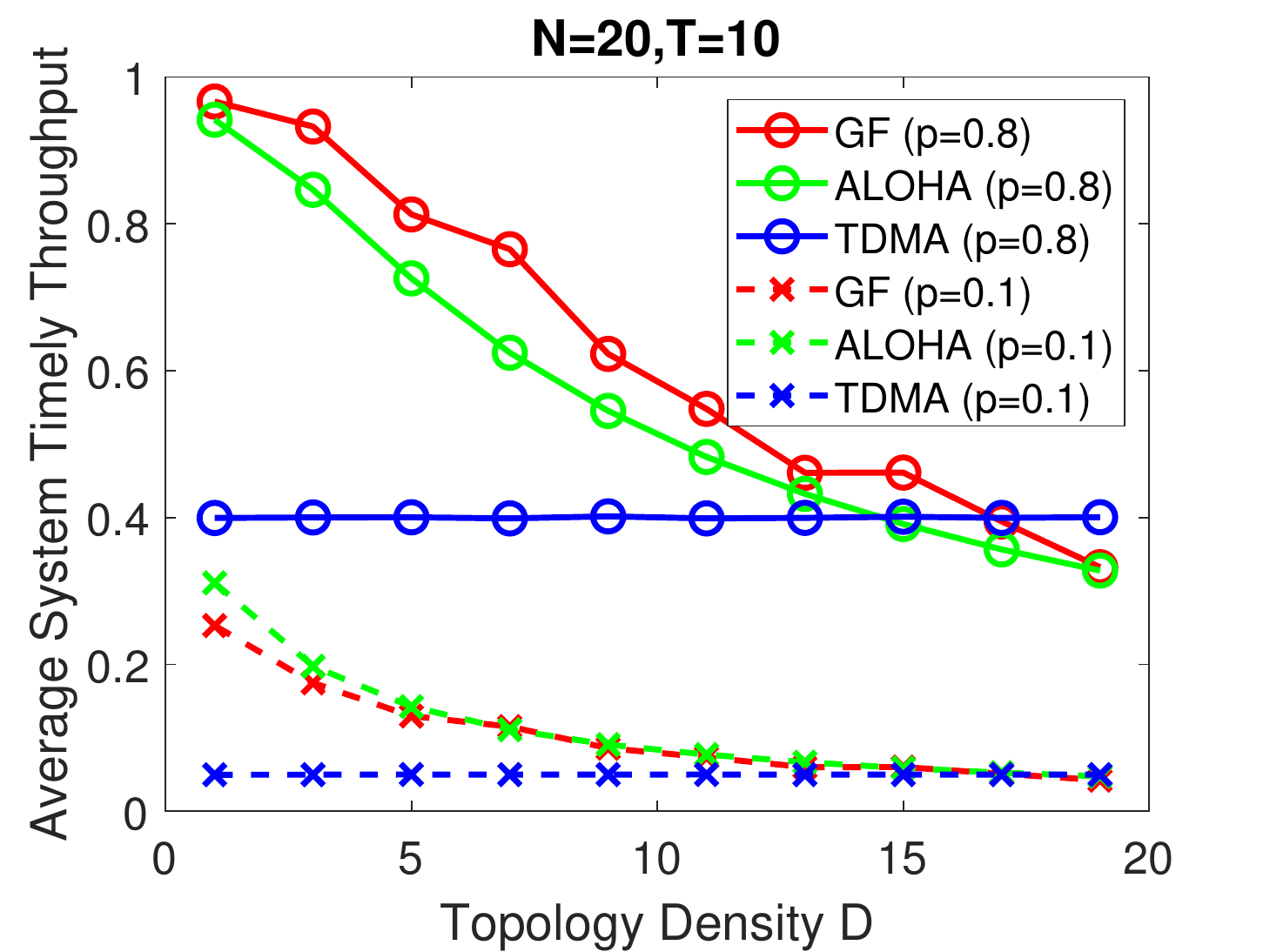}
  \caption{{The performance of the proposed GF, ALOHA, and TDMA schemes under a poisson-arrival traffic pattern where the mean of the inter-arrival time is $10$.}}\label{fig:effect-poisson}
\end{figure}

\subsection{Benefit of Feedback Information} \label{subsec:sim-feedback}
In this paper, we assume that there is no feedback information from receivers to transmitters. If feedback is available,
we can still apply the proposed GF, ALOHA, and TDMA schemes. However, since each transmitter can get the feedback from the receiver about whether
its packet has been delivered successfully or not, it can terminate all transmissions in the rest of a frame after a successful delivery.
This can reduce the competition and thus potentially increase the system performance.
{We consider an example with $N=20, T=30, p_i=p=0.8\; (\forall i)$ and varying $D$.
For each $D$, we randomly generate 100 topologies and calculate the mean values of average system timely throughput for the 100 topologies with and without feedback information. Fig.~\ref{fig:feedback-gain} shows the feedback gain (the difference between throughput with feedback and throughput without feedback)  for GF, ALOHA and TDMA schemes.
}
As we can see, the feedback information provides more gains for ALOHA than GF, while TDMA has no feedback gain because TDMA scheme does not introduce competition.

\begin{figure}
  \centering
  \includegraphics[width=0.8\linewidth]{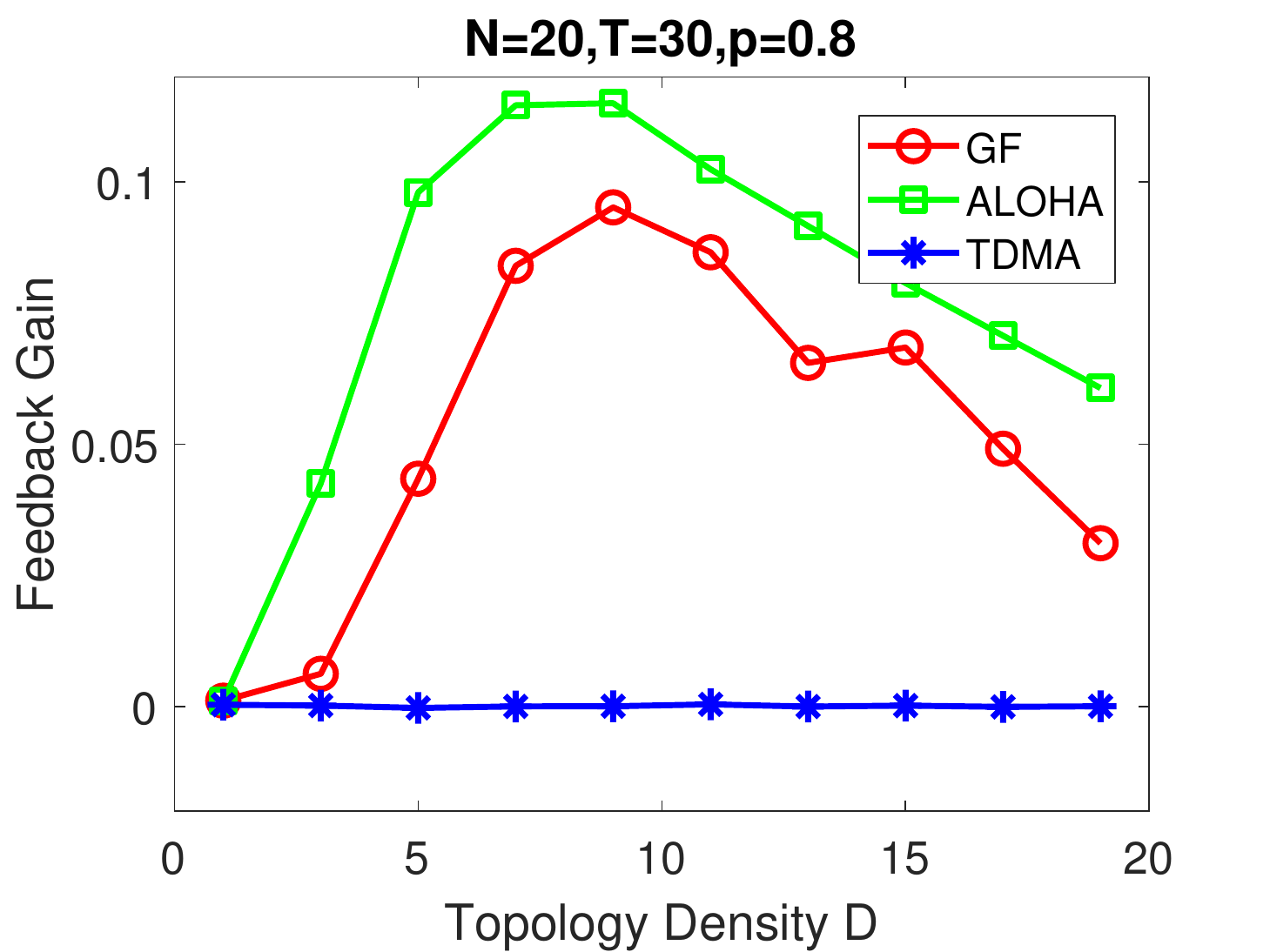}
  \caption{{Feedback gain of GF, ALOHA, and TDMA schemes.}}\label{fig:feedback-gain}
\end{figure}

\subsection{Performance in a Practical MANET Environment} \label{subsec:MANET}
In this paper, we have adopted several modeling simplifications such that we can compare different topology-transparent distributed schemes analytically.
It is important to further evaluate the performance of different schemes in a practical MANET environment.
In this subsection, we follow \cite{lutz2017variable} to simulate a practical MANET environment.
The environment consists of 50 nodes moving at the speed of 30m/s. The nodes are equipped with omni-directional antennas with hearing distance being 200m.
The duration of each slot is 0.8ms. Both sparse distribution and dense distribution are investigated. In the sparse distribution case, the 50 nodes are randomly placed in a 500m by 500m area
with an expected network density $D=27$.
In the dense distribution case, the 50 nodes are randomly placed in a 2000m by 2000m area with an expected network density $D=4$.

In addition, we adopt the physical wireless channel model in \cite{gong2020age} to calculate wireless channel quality $p_i$, which is a key modeling parameter in our analysis. In the model, for collision-free transmission from a transmitter to a receiver, the small-scale fading  follows an exponential random variable with a unit mean. The large-scale fading is denoted by $d_i^{\tau}$, where $d_i$ represents the distance between the transmitter and the receiver of pair $i$, and $\tau$ is the path-loss factor. The additive white Gaussian noise follows a complex Gaussian distribution $N(0,\delta^2)$. According to \cite{gong2020age}, the probability that a collision-free packet of pair $i$ can be successfully delivered at a time slot is
\begin{equation} \label{eq: physical-prob}
p_i =\exp\left(-\dfrac{\delta^2 d_i^{\tau} (2^{r_{th}}-1)}{P} \right),
\end{equation}
where $P$ is the transmission power of the transmitter and $r_{th}$ is the threshold value for transmission rate.
It is obvious that larger transmission power $P$ corresponds to larger channel quality $p_i$ according to \eqref{eq: physical-prob}.
We simulate throughput performance for this model with different $P$ from 0.1W to 0.5W. The distance $d_i$ is assumed to follow a uniform distribution from 50m to 150m, $\forall i\in \{1,2,\ldots,N\}$. We set $\tau=3$, $\delta^2=10^{-7}$W/Hz and $r_{th} = 1$bps/Hz.

The simulation parameters of our considered MANET environment are summarized in Table~\ref{tab:summary of parameters}.

\begin{table}[t]
\caption{{Summary of simulation parameters. \label{tab:summary of parameters}}}
\centering
\begin{tabular}{|c|c|}
\hline
Parameter        & Value          \\ \hline
Number of nodes $N$           & 50 \\ \hline
Moving speed            & 30m/s                     \\ \hline
Hearing distance $\Delta$             & 200m \\ \hline
Slot duration              & 0.8ms           \\ \hline
Sparse distribution area & 2000m by 2000m \\ \hline
Dense distribution area & 2000m by 2000m \\ \hline
Distance between transmitter-receiver pair $d_i$ & $U(50\text{m},150\text{m})$ \\ \hline
Path loss factor $\tau$ & 3 \\ \hline
Noise power density $\delta^2$  & $10^{-7}$W/Hz \\ \hline
Transmission rate  threshold $r_{th}$ & 1bps/Hz \\ \hline
Transmission power levels & 0.1W-0.5W \\ \hline
\end{tabular}
\end{table}

\begin{figure}[t]
  \centering
  \subfigure[Sparse distribution.]{
    \label{fig:effect-power1} 
    \includegraphics[width=0.8\linewidth]{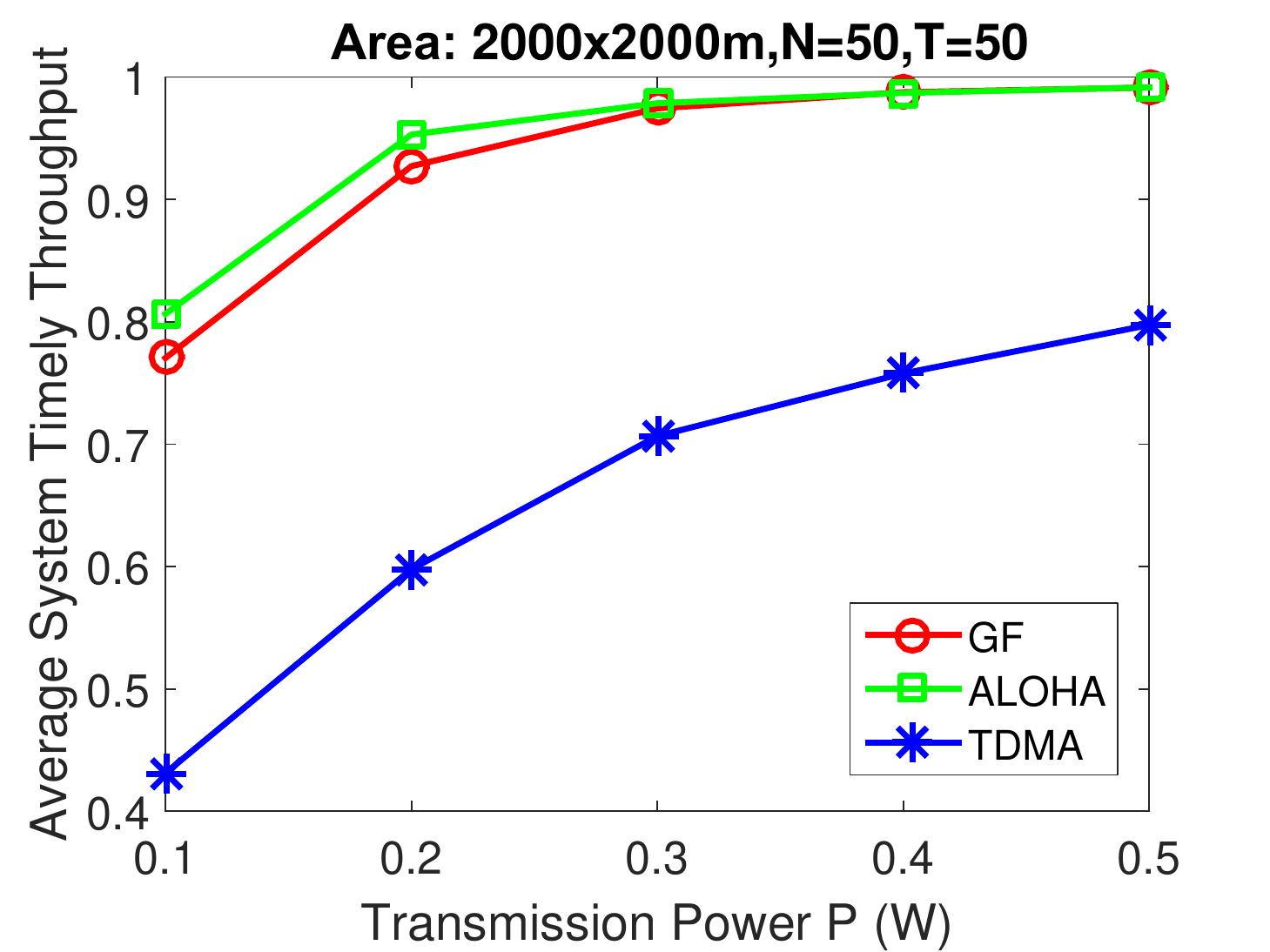}}
    \\ 
    \subfigure[Dense distribution.]{
    \label{fig:effect-power2} 
    \includegraphics[width=0.8\linewidth]{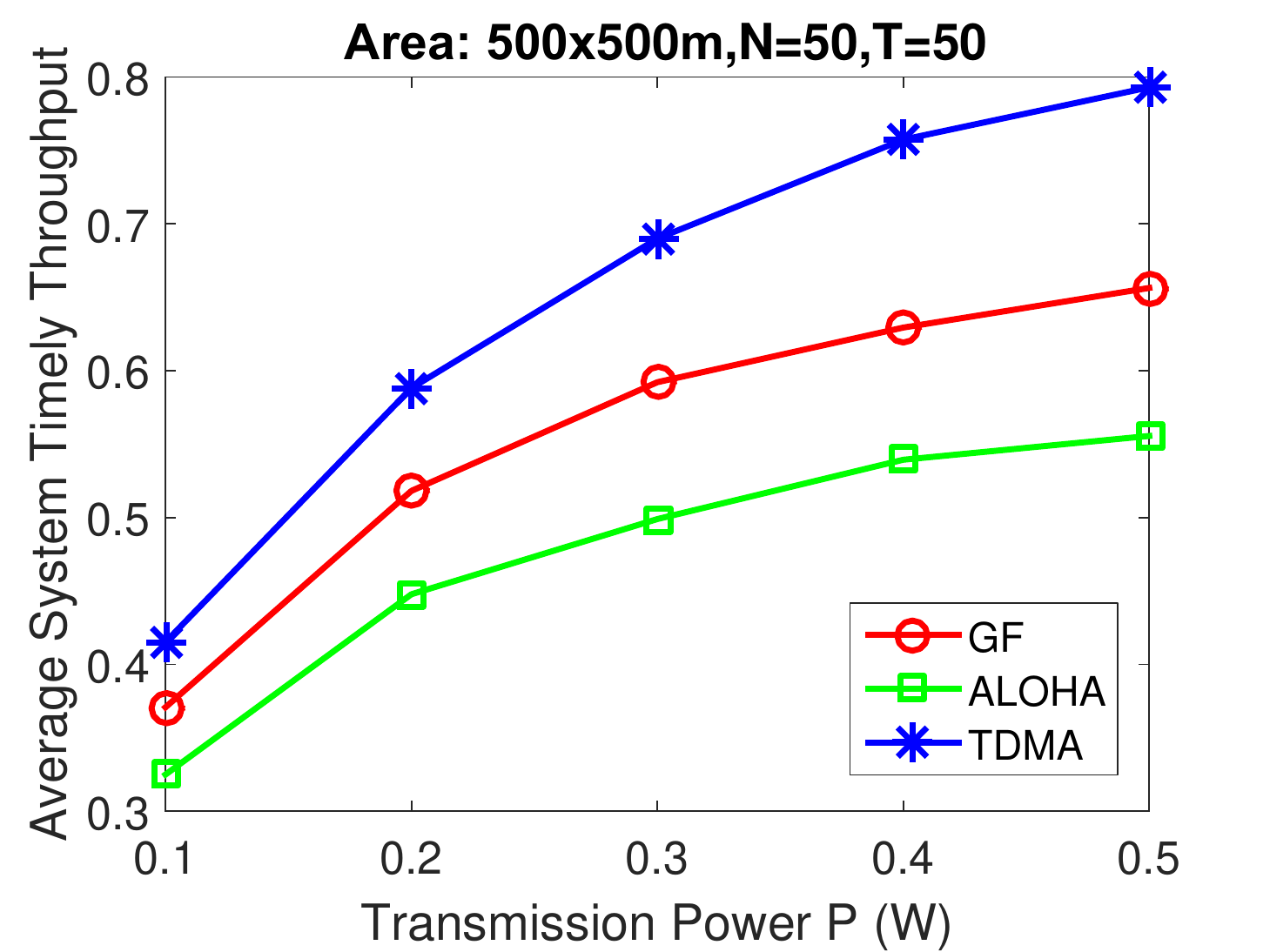}}
  \caption{{Effect of transmission power under a practical MANET environment in Sec.~\ref{subsec:MANET}.} \label{fig:effect-power} } 
\end{figure}

We plot the throughput performance of ALOHA, TDMA and the GF sequence scheme with different power levels for both spare and dense scenarios in Fig.~\ref{fig:effect-power}.
From Fig.~\ref{fig:effect-power}, we can observe that the throughput performance under GF and ALOHA is better than that under TDMA in the sparse scenario (i.e., when $D$ is small), and TDMA achieves the largest throughput in the dense scenario (i.e., when $D$ is large). For the small $D$ case, when the transmission power $P$ is small, which means that the channel quality is low, we can observe from Fig.~\ref{fig:effect-power1} that  ALOHA outperforms GF. Thus, even under our considered practical MANET environment,
the results are in line with the those in Fig.~\ref{fig:effect-p} where we adopt several modeling simplifications, including the simplified $p_i$-parameterized wireless channel model.
}
\fi


\begin{table}[t]
\caption{Summary of the best settings for different schemes. \label{tab:summary}}
\centering
\begin{tabular}{|c|c|}
\hline
\textbf{Schemes}          & \textbf{Best Settings}             \\ \hline
ALOHA            & \specialcell{Network density $D$ is small  \\ and channel quality $\{p_i\}$ is small}  \\ \hline
TDMA             & Network density $D$ is large                     \\ \hline
The GF Sequence Scheme               & \specialcell{ Network density $D$ is small \\ and channel quality $\{p_i\}$ is large }             \\ \hline
The Combination Sequence Scheme              & Network density $D=1$           \\ \hline
\end{tabular}
\end{table}

\section{Conclusion} \label{sec:conclusion}
In this paper, distributed
scheduling designs for a topology-transparent MANET to support delay-constrained
traffic are investigated for the first time. We have analyzed and compared the average system timely throughput
of several  schemes including ALOHA, TDMA, the GF sequence scheme and the combination sequence scheme.
Different schemes work best for different settings.
We have summarized their individual best settings in Table~\ref{tab:summary} according
to our analysis and simulations in this paper.

Our main contribution in this work is that we have analyzed and compared different  distributed schemes which
were generally designed for delay-unconstrained setting. In the future, it would be interesting
to design a novel distributed scheme that is particularly suitable for delay-constrained
setting. Here we give some of our thoughts on how to design such a new scheme.
One direction is to design a \emph{hybrid} scheme combining both the deterministic sequence scheme and the probabilistic scheme.
The sequence scheme utilizes the network topology elegantly such that each user has at least one collision-free slot in a period.
However, once two or more pairs have bit `1' in the same slot, they will collide for sure. The probabilistic scheme can soften the collision such that
a user can still have chance to deliver its packet successfully even other users also have bit `1' in the same slot.
It is possible to combine the benefits of both the deterministic sequence scheme and the probabilistic scheme to design a better hybrid scheme.
Another direction is to re-design the sequence assignment mechanism. In our paper, we assume that sequences are pre-assigned to the users
and a user will keep using its assigned sequence all the time. This approach looks inflexible.
It is possible to design a sequence pool from which each user can randomly select a sequence \cite{chang2019asynchronous}. Users can also adaptively change
the sequences. This approach increases the flexibility. The research problem is how to design a good sequence pool.
We will work along these directions in the future.

\bibliographystyle{IEEEtran}
\bibliography{ref}

\ifx \ISTR \undefined
\else

\begin{appendix}

\subsection{Proof of Theorem~\ref{thm:theory-R-i-L-T}} \label{app:proof-of-lemma-theory-R-i-L-T}
\subsubsection{Proof of Case 1}
When $L \ge T$, as exemplified in Fig.~\ref{fig:case 1},
we consider the first super period from period 1 to period $T$ without loss of generality.
In those slots, transmitter $ i $ has $T$ collision-free slots.
In addition, since the distance of any two consecutive collision-free slots, which is $L \ge T$,
is large than the distance between the beginning of a frame and the end of the frame, which is $T-1$,
any two consecutive collision-free slots cannot belong to the same frame.
Therefore, all $T$ collision-free slots belong to $ T $ different frames.
For example, in Fig.~\ref{fig:case 1} with $L=4$, $T=3$ and offset $t_k=2$,
we can see that transmitter $i$ has $T=3$ collision-free slots distributed in  $T=3$ different frames.
Thus, in this super period, $T$ packets can be transmitted once, each of which can be
delivered successfully with  probability $p_i$.
Hence, the average number of packets delivered before expiration in this super period is $ Tp_i $,
which holds for any super period that satisfies the location-fixed condition.
Thus, the timely throughput of pair $i$ is
\be
R^{\textsf{case-1}}_i = \frac{ \mathbb{E} \left[  \substack{\text{number of packets delivered before } \\ \text{expiration from slot 1 to slot $LT$}} \right]}{LT/T}
= \frac{Tp_i}{LT/T} = \frac{Tp_i}{L}, \nnb
\ee
which completes  the proof of Case 1 in Theorem~\ref{thm:theory-R-i-L-T}.

\subsubsection{Proof of Case 2}
When $L < T$, as exemplified in Fig.~\ref{fig:case 2},
we again consider the first super period from period 1 to period $T$ without loss of generality.
In addition, we assume that the offset is $t_k=1$. The proof can be easily extended to general offset $t_k \in \{1,2,\cdots,L\}$.
Next we will show that each of the $L$ frame in this super period is either a \emph{Type-1 frame}
where there are  $ \left \lceil T/L \right \rceil  $ collision-free slots or a \emph{Type-2 frame}
where there are  $ \left \lfloor T/L \right \rfloor  $ collision-free slots.\footnote{
Our proofs in the rest of this part still hold when $T \bmod L = 0$, under which
the Type-1 frame and Type-2 frame have no difference.} In addition,
there are $\alpha = (T \bmod L)$ Type-1 frames and $\beta = L - \alpha$ Type-2 frames in this super period.
For example, in Fig.~\ref{fig:case 2} with $L=3$ and $T=4$,
we can see that there is $\alpha = (T \bmod L) = 1$ Type-1 frame which has
$ \left \lceil T/L \right \rceil  = 2$ collision-free slots and there are $\beta = L -\alpha = 2$
Type-2 frames each of which has  $ \left \lfloor T/L \right \rfloor  = 1$ collision-free slot.

%

Clearly, since the distance of any two consecutive collision-free slots, i.e., $L$,
is less than the frame length $T$,
any frame has at least one collision-free slot. Then,
for any frame $i \in \{1,2,\cdots,L\}$, we define its \emph{one-offset $d_i$} as
the distance between its first slot and its first collision-free slot after the first slot.
For example, in Fig.~\ref{fig:case 2} with $L=3$ and $T=4$, the one-offsets of frames 1, 2, and 3 are 3, 2, and 1 respectively.
Since there is a collision-free slot every $L$ slots,
we can see that $d_i \in \{1,2,\cdots,L\}$.
When the one-offset of frame $i$ is $d_i=L$,
the first slot of frame $i$ is a collision-free slot and there are $\lceil \frac{T}{L} \rceil$ collision-free slots in frame $i$,
implying that frame $i$ is a Type-1 frame.
When $d_i \in \{0,1,\cdots,L-1\}$, the first slot of frame $i$ is not a collision-free slot
and there are $\lceil \frac{T-d_i}{L} \rceil$ collision-free slots in frame $i$.
It is straightforward to show that frame $i$ is a Type-1 frame if its one-offset
\be
d_i\in  \mathcal{D}_1 \triangleq \{1,2,\cdots, (T \bmod L)-1 \} \cup \{L\},
\label{equ:def-D-1}
\ee
and it is a Type-2 frame if its one-offset
\be
d_i\in  \mathcal{D}_2 \triangleq  \{T \bmod L, (T \bmod L) + 1, \cdots,  L-1 \}.
\label{equ:def-D-2}
\ee

\begin{figure}[t]
    \centering
        \includegraphics[width=0.5\linewidth]{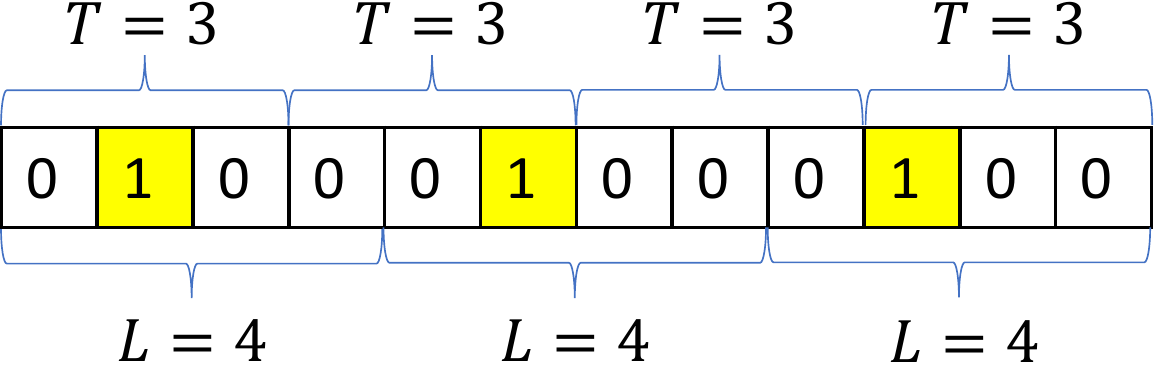}
    \caption{Illustration for the case $L \ge T$ where ``1" indicates a collision-free slot. Here the offset is $t_k=2$.
    We can see that transmitter $i$ has $T=3$ collision-free slots distributed in  $T=3$ different frames.
    } \label{fig:case 1} \vspace{-0.2cm}
\end{figure}

\begin{figure}[t]
    \centering
        \includegraphics[width=0.5\linewidth]{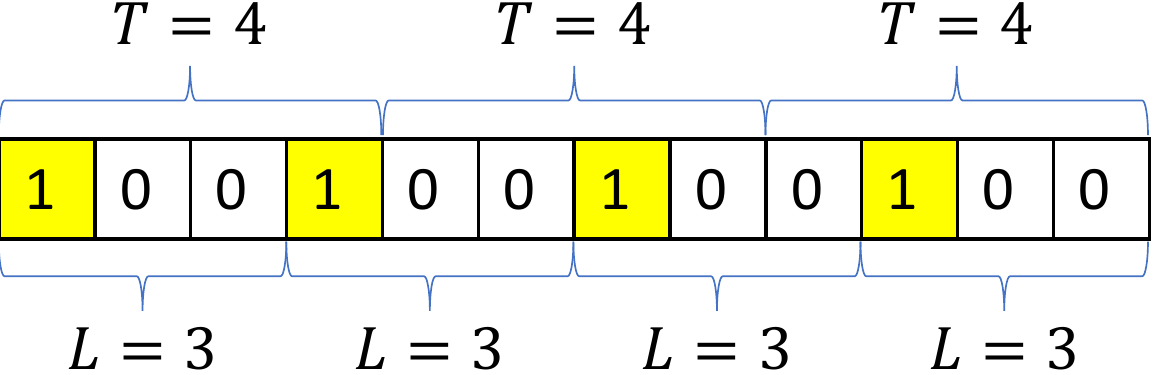}
    \caption{Illustration for the case $L < T$ where ``1" indicates a collision-free slot. Here the offset is $t_k=1$.
    We can see that there is $\alpha = (T \bmod L) = 1$ Type-1 frame which has
    $ \left \lceil T/L \right \rceil  = 2$ collision-free slots and there are $\beta = L -\alpha = 2$
    Type-2 frames each of which has  $ \left \lfloor T/L \right \rfloor  = 1$ collision-free slot.} \label{fig:case 2} \vspace{-0.5cm}
\end{figure}

Denote the greatest common divisor of $T$ and $L$ by $s$ and define positive integers $ T'=T/s $, $L'=L/s$. Note that $T'$ and $L'$ are coprime integers.
We have
\be
T \bmod L = s(T' \bmod L').
\ee
Note that both the traffic pattern and the sequence are completely the same every $L'$ frames in the first super period,
which contains $L/L'=s$ copies consisting of such $L'$ frames.
We then consider the first $L'$ frames from slot 1 to slot $L'T$.
The one-offset of frame $ i\in \{1,2,\ldots, L' \} $ is
\bee
d_i & =L-((i-1)T \bmod L)=sL'-s((i-1)T' \bmod L') \nnb \\
& = s[L' - ((i-1)T' \bmod L')]. \nnb
\eee
Since $((i-1)T' \bmod L') \in \{0,1,\cdots, L'-1\}$ for any $i \in \{1,2,\cdots, L'\}$, we have
$[L' - ((i-1)T' \bmod L')] \in \{1,2,\cdots,L'\}$ and $d_i=s[L' - ((i-1)T' \bmod L')] \in \Omega \triangleq \{s,2s, \cdots, L's\}$.
Next we use contradiction to prove that
\be
\{d_i: i=1,2,\cdots,L'\} = \Omega.
\label{equ:d-i-is-equal-to-a-set}
\ee
Suppose not. Namely, there exist $ i_1, i_2 \in \{1,2,\ldots, L' \} , i_1\neq i_2 $ such that $ d_{i_1}=d_{i_2} $, i.e.,
\be
(i_1-1) T' \equiv (i_2-1) T' \bmod L',
\ee
implying
\be \label{e1-appendix}
 (i_1-i_2)T' \equiv 0 \bmod L'.
\ee
Since $-(L'-1)\leq i_1-i_2\leq (L'-1)$ and $ T' $ is coprime with $L'$, \eqref{e1-appendix} cannot hold.
This is a contradiction. Therefore, \eqref{equ:d-i-is-equal-to-a-set} holds.
Note that
\bee
& \Omega  = \{s,2s, \cdots, L's\}  \nnb \\
&    \resizebox{.99\linewidth}{!}{$=\{s, 2s, \cdots, s(T' \bmod L'), s(T' \bmod L' +1), \cdots, s(L'-1), sL'\}$}  \nnb \\
& \resizebox{.99\linewidth}{!}{$=  \{\underbrace{s, 2s, \cdots}_{\mathcal{D}_1}, \underbrace{T \bmod L, s(T' \bmod L' +1), \cdots, s(L'-1)}_{\mathcal{D}_2}, \underbrace{L}_{\mathcal{D}_1}\}.$}  \nnb
\eee
Thus, in the first $L'$ frames, the number of Type-1 frames is $(T' \bmod L')$ and the number of Type-2 frames is $[L' - (T' \bmod L')]$.
In the first super period which contains $s$ copies of such $L'$ frames,
the number of Type-1 frames is $s(T' \bmod L')=T \bmod L = \alpha$ and the number of Type-2 frames is $s[L' - (T' \bmod L')]=L - (T \bmod L) = L - \alpha = \beta$.

In a Type-1 (resp. Type-2) frame with $ \left \lceil T/L \right \rceil  $ (resp. $ \left \lfloor T/L \right \rfloor$) collision-free slots,
the packet will be transmitted $ \left \lceil T/L \right \rceil  $  (resp. $ \left \lfloor T/L \right \rfloor$) times and it will be
delivered successfully with probability  $ 1-(1-p_{i})^{ \left \lceil \frac{T}{L} \right \rceil }$ (resp. $ 1-(1-p_{i})^{ \left \lfloor \frac{T}{L} \right \rfloor }$).
Hence, the average number of packets delivered before expiration in the first super period is
\bee
& \alpha  \left[1-(1-p_{i})^{ \left \lceil \frac{T}{L} \right \rceil } \right] + \beta \left[1-(1-p_{i})^{ \left \lfloor \frac{T}{L} \right \rfloor }\right],
\eee
and the timely throughput of pair $i$ is
\bee
& R^{\textsf{case-2}}_i = \frac{ \mathbb{E} \left[  \substack{\text{number of packets delivered before } \\ \text{expiration from slot 1 to slot $LT$}} \right]}{LT/T} \nnb \\
&  =\frac{  \alpha \left[1-(1-p_{i})^{ \left \lceil \frac{T}{L} \right \rceil} \right]+\beta \left[1-(1-p_{i})^{ \left \lfloor \frac{T}{L} \right \rfloor} \right]}{L},
\eee
where $\alpha = (T \bmod L)$ and $\beta = L - \alpha$.
This completes the proof of Case 2 in Theorem~\ref{thm:theory-R-i-L-T}.

\subsection{Proof of Lemma~\ref{lem:R-1-R-2-decrease-with-L}} \label{app:proof-of-lem-R-1-R-2-decrease-with-L}

Clearly $R_i^{\textsf{case-1}}(L,T)$ in \eqref{equ:avg-1} strictly decreases as $L$ increases.

When $p_i=1$, we can see that $R_i^{\textsf{case-2}}(L,T)=1$ in \eqref{equ:avg-2}. Then we only need to
to show that $R_i^{\textsf{case-2}}(L,T)$ in \eqref{equ:avg-2} strictly decreases as $L$ increases when $p_i \in (0,1)$.
We first note that\footnote{We denote $\alpha(L,T) = \alpha = T \bmod L$ to describe explicitly the dependence of $\alpha$ on $L$ and $T$.}
\bee
& R_i^{\textsf{case-2}}(L,T)  = \frac{  \alpha \left[1-(1-p_{i})^{ \left \lceil \frac{T}{L} \right \rceil} \right]+\beta \left[1-(1-p_{i})^{ \left \lfloor \frac{T}{L} \right \rfloor} \right] }{L} \nnb \\
& = \frac{\alpha}{L} \left[1-(1-p_{i})^{ \left \lceil \frac{T}{L} \right \rceil} \right]+ \left(1- \frac{\alpha}{L}\right)  \left[1-(1-p_{i})^{ \left \lfloor \frac{T}{L} \right \rfloor} \right] \nnb \\
& = \frac{\alpha(L,T)}{L} f_1(L,T) + \left(1- \frac{\alpha(L,T)}{L}\right) f_2 (L,T), \nnb
\eee
is a convex combination of $f_1(L,T) \triangleq 1-(1-p_{i})^{ \left \lceil \frac{T}{L} \right \rceil}$
and $f_2(L,T) \triangleq 1-(1-p_{i})^{ \left \lfloor \frac{T}{L} \right \rfloor}.$

When $ \left \lceil \frac{T}{L} \right \rceil =  \left \lfloor \frac{T}{L} \right \rfloor$, we have
$f_1(L,T)=f_2(L,T)$ and
\[
R_i^{\textsf{case-2}}(L,T) = f_1(L,T) = 1-(1-p_{i})^{ \left \lceil \frac{T}{L} \right \rceil}
\]
which strictly decreases as $L$ increases.

When $ \left \lceil \frac{T}{L} \right \rceil \neq  \left \lfloor \frac{T}{L} \right \rfloor$, we have
$f_2(L,T) < f_1(L,T)$. Since $\alpha(L,T)= (T \bmod L) \in \{0,1,\cdots, L-1\}$, we have
$\frac{\alpha(L,T)}{L} \in [0,1).$
Thus, we have
\be
f_2(L,T) \le R_i^{\textsf{case-2}}(L,T) < f_1(L,T).
\label{equ:app-f2-Ri2-f1}
\ee

To prove that $ R_i^{\textsf{case-2}}(L,T)$ in \eqref{equ:avg-2} strictly decreases as $L$ increases, we only need to show that
\be
R_i^{\textsf{case-2}}(L,T) > R_i^{\textsf{case-2}}(L+1,T), \quad \forall 1 \le L \le T-1.
\label{equ:app-Ri2-L-T-larger-than-Ri2-L+1-T}
\ee

Toward that end, we use the following inequality,
\be
\left \lceil \frac{T}{L+1} \right \rceil \le \left \lceil \frac{T}{L} \right \rceil \le \left \lfloor \frac{T}{L} \right \rfloor + 1.
\label{equ:app-a-key-inequ-ceil-and-floor}
\ee
We then consider two cases.

\emph{Case I. $\left \lceil \frac{T}{L+1} \right \rceil \le  \left \lfloor \frac{T}{L} \right \rfloor$}. In this case, based on \eqref{equ:app-f2-Ri2-f1},
we have
\bee
& R_i^{\textsf{case-2}}(L+1,T) < f_1(L+1,T)  = 1-(1-p_{i})^{ \left \lceil \frac{T}{L+1} \right \rceil} \nnb \\
& \le 1-(1-p_{i})^{ \left \lfloor \frac{T}{L} \right \rfloor} = f_2(L,T) \le R_i^{\textsf{case-2}}(L,T).
\eee
Thus, we have proved \eqref{equ:app-Ri2-L-T-larger-than-Ri2-L+1-T} and therefore
$R_i^{\textsf{case-2}}(L,T)$ in \eqref{equ:avg-2} strictly decreases as $L$ increases in this case.

\emph{Case II. $\left \lceil \frac{T}{L+1} \right \rceil =  \left \lfloor \frac{T}{L} \right \rfloor+1$}.
In this case, both inequalities in \eqref{equ:app-a-key-inequ-ceil-and-floor} hold as equalities, i.e.,
\be
\left \lceil \frac{T}{L+1} \right \rceil = \left \lceil \frac{T}{L} \right \rceil = \left \lfloor \frac{T}{L} \right \rfloor + 1.
\label{equ:app-a-key-equ-ceil-and-floor}
\ee
The second equality in \eqref{equ:app-a-key-equ-ceil-and-floor}, i.e, $\left \lceil \frac{T}{L} \right \rceil = \left \lfloor \frac{T}{L} \right \rfloor + 1$,
 implies that $T \bmod L \neq 0$.
In addition, the whole equality in \eqref{equ:app-a-key-equ-ceil-and-floor}, i.e., $\left \lceil \frac{T}{L+1} \right \rceil=\left \lfloor \frac{T}{L} \right \rfloor + 1$,
implies that
\be
 \resizebox{.892\linewidth}{!}{$
\left \lfloor  \frac{T}{L} \right \rfloor = \left \lceil \frac{T}{L+1} \right \rceil - 1
 < \frac{T}{L+1} < \frac{T}{L} < \left \lceil \frac{T}{L} \right \rceil = \left \lfloor  \frac{T}{L} \right \rfloor  + 1,$}
\ee
i.e.,
\be
\left \lfloor  \frac{T}{L} \right \rfloor  < \frac{T}{L+1} <  \left \lfloor  \frac{T}{L} \right \rfloor + 1.
\ee

Therefore, we must have
\be
\left \lfloor \frac{T}{L+1} \right \rfloor = \left \lfloor \frac{T}{L} \right \rfloor,  \left \lceil \frac{T}{L+1} \right \rceil = \left \lceil \frac{T}{L} \right \rceil.
\ee

Since both $\frac{T}{L}$ and $\frac{T}{L+1}$ are not integers and have the same floor and ceiling, we have
\be
 \resizebox{.892\linewidth}{!}{$
\alpha(L+1,T) = T \bmod (L+1) = (T \bmod L) - 1 = \alpha(L,T) - 1,$}
\label{equ:app-alpha-T+1-alpha-T}
\ee
and
\be
f_2(L,T) = f_2(L+1,T) <  f_1(L+1,T) = f_1(L,T).
\label{equ:app-f1-T+1-f1-T-f2-T+1-f2-1}
\ee

Therefore, based on \eqref{equ:app-alpha-T+1-alpha-T} and \eqref{equ:app-f1-T+1-f1-T-f2-T+1-f2-1}, we have
\bee
& R_i^{\textsf{case-2}}(L,T) - R_i^{\textsf{case-2}}(L+1,T)  \nnb \\
& = \frac{\alpha(L,T)}{L} f_1(L,T) + \left(1- \frac{\alpha(L,T)}{L}\right) f_2 (L,T) - \nnb \\
&  \frac{\alpha(L+1,T)}{L+1} f_1(L+1,T) - \left(1- \frac{\alpha(L+1,T)}{L+1}\right) f_2 (L+1,T) \nnb \\
& = \frac{\alpha(L,T)}{L} f_1(L,T) + \left(1- \frac{\alpha(L,T)}{L}\right) f_2 (L,T) \nnb \\
& - \frac{\alpha(L,T)-1}{L+1} f_1(L,T) - \left(1- \frac{\alpha(L,T)-1}{L+1}\right) f_2 (L,T) \nnb \\
& = \frac{(L+\alpha(L,T))(f_1(L,T)-f_2(L,T))}{L(L+1)} > 0.
\eee
Thus, we have proved \eqref{equ:app-Ri2-L-T-larger-than-Ri2-L+1-T} and therefore
$R_i^{\textsf{case-2}}(L,T)$ in \eqref{equ:avg-2} strictly decreases as $L$ increases in this case.

\emph{Case I} and \emph{Case II} complete the proof.

\subsection{Proof of Proposition~\ref{prop:min-L-for-D=1} } \label{proof-of-thm-min-L-for-D=1}
To prove this proposition, we consider another related problem:
what is the maximum number of pairs (i.e., $N$), denoted as $N^{\max}(D,L)$,  under the requirement that we can assign each pair a unique sequence of period $L$
which has at least one collision-free slot in a period for any network topology with density $D$?
It turns out that this problem is  to find the maximum set size of \emph{D-cover-free families} \cite{furedi1996onr}. For the special case of $D=1$ where any sequence is not blocked by any other sequence,
the problem can be answered according to Sperner's theorem \cite[Theorem 1.2.1]{anderson1987combinatorics},
\[
N^{\max}(1,L) = \binom{L}{\left\lceil \frac{L}{2} \right\rceil}.
\]
Therefore, the minimal sequence period $L$ to support a set of $N$ sequences each of which has at least one collision-free slot
is
\[
L^{\min} (1,N) = \min \left \{L \in \mathbb{Z}^+: \binom{L}{\left\lceil \frac{L}{2} \right\rceil} \ge N \right \},
\]
which completes the proof.

\subsection{Proof of Theorem~\ref{thm:aloha-decreasing-with-D}} \label{app:proof-of-thm-aloha-decreasing-with-D}
Let $f(D)=\dfrac{1}{D+1} \left(1-\dfrac{1}{D+1}\right)^D$. Then we have
\[
\underline{R}^{\textsf{ALOHA}^*}(D,N,T)=\dfrac{\sum_{i=1}^N \left\{1-\Pi_{i=1}^T \left[1-f(D)p_i \right] \right \}}{N}.
\]
We can observe that for any fixed $ N $, $T$ and $\{p_i\}$, if $ f(D) $ is strictly decreasing with respect to $D$, then $ \underline{R}^{\textsf{ALOHA}^*}(D,N,T) $ is also strictly decreasing with respect to $D$. Thus, we only need to prove that $ f(D) $ is strictly decreasing with respect to $D$. Note that
\bee
\ln f(D) &= \ln\dfrac{1}{D+1}+D\ln\left(1-\dfrac{1}{D+1}\right) \nnb \\
&=D\ln D -(D+1)\ln (D+1).
\label{equ:ln-f-D}
\eee
Taking derivative with respect to $D$ in both sides of \eqref{equ:ln-f-D}, we obtain that
\[
f'(D)=f(D) \left[\ln D - \ln (D+1) \right] <0.
\]
Thus $\underline{R}^{\textsf{ALOHA}^*}$ is strictly decreasing with respect to $D$,
which completes the proof.

\subsection{Proof of Proposition~\ref{prop:D-N-1-TDMA-better-than-GF}} \label{app:proof-of-lem-D-N-1-TDMA-better-than-GF}
The sequence period of TDMA is $N$. When $D = N-1$ and $N$ is a prime power, the smallest prime power $q$ that
satisfies \eqref{equ:min-q-GF} is $q(D,N)=q(N-1,N)=N$. Thus, the sequence period of the GF sequence scheme is $q^2(N-1,N)=N^2$.

We consider a \emph{super TDMA sequence} of length $N^2$, which repeats a TDMA sequence (of length $N$) $N$ times.
Then a super TDMA sequence has $N$ collision-free slots, which are distributed uniformly in the sense that any two adjacent
collision-free slots are in the distance of $N$ slots. However, a GF sequence has exactly $q(N-1,N)=N$ ones and thus
has at most $N$ collision-free slots, which are not necessarily distributed uniformly.
Our proof can be divided into two cases: $T\le N$ and $T>N$. In both of the two cases, we consider a super frame of $N^2T$ slots,
which has in total $N^2$ frames/packets, in total $NT$ collision-free slots under TDMA scheme, and at most $NT$ collision-free slots under the GF sequence scheme.

\emph{Case I: $T \le N$.} 
In this case,  under the  TDMA scheme, similar to the analysis in the proof of Case I in Theorem~\ref{thm:theory-R-i-L-T}, all $NT$ uniformly-distributed collision-free slots belong
to different frames. The expected number of delivered packets before expiration in a super frame of $N^2T$ slots is $NTp_i$.
On the other hand, under the GF sequence, there are at most $NT$ collision-free slots in a super frame of $N^2T$ slots, but
some collision-free slots could belong to the same frames. Suppose that the total number of frames that all  collision-free slots occupy is $K$
and the $k$-th such frame is occupied by $n_k \in \mathbb{Z}^+$ collision-free slots. Thus, we have
\be
n_1 + n_2 + \cdots + n_K \le NT, \nnb
\ee
and the expected number of delivered packets before expiration in the super frame is
\be
\left[1-(1-p_i)^{n_1}\right] + \left[1-(1-p_i)^{n_2}\right] + \cdots + \left[1-(1-p_i)^{n_K}\right]. \nnb
\ee
It is easy to use induction to prove the following inequality
\be
1-(1-p_i)^n \le np_i, \quad \forall n \in \mathbb{Z}^+. \nnb
\ee
Thus, we have
\bee
& \left[1-(1-p_i)^{n_1}\right] + \left[1-(1-p_i)^{n_2}\right] + \cdots + \left[1-(1-p_i)^{n_K}\right] \nnb \\
& \le n_1 p_i + n_2 p_i + \cdots + n_K p_i = p_i \sum_{k=1}^{K}n_k  \le NT p_i. \nnb
\eee
Therefore, the expected number of delivered packets before expiration in a super frame under the GF sequence scheme
is not greater than that under the TDMA scheme. Thus, TDMA achieves larger (or equal) average system timely throughput.

\begin{table*}[t]
\centering
\caption{$q(1,N)$ with different $N$. \label{tab:q-1-N}}
\begin{tabular}{|l|l|l|l|l|l|l|l|l|l|l|l|l|l|l|l|l|l|l|l|}
\hline
$N$             & 1  & 2  & 3  & 4  & 5  & 6  & 7  & 8  & 9  & 10  & 20  & 30  & 40  & 50  & 60  & 70  & 80  & 90  & 100 \\ \hline
$q(1,N)$        & 2  & 2  & 2  & 2  & 3  & 3  & 3  & 3  & 3  & 3   & 3   & 4   & 4   & 4   & 4   & 4   & 4   & 4   & 4   \\ \hline
$q^2(1,N) $     & 4  & 4  & 4  & 4  & 9  & 9  & 9  & 9  & 9  & 9   & 9   & 16  & 16  & 16  & 16  & 16  & 16  & 16  & 16  \\ \hline
$q^2(1,N) < N?$ & No & No & No & No & No & No & No & No & No & Yes & Yes & Yes & Yes & Yes & Yes & Yes & Yes & Yes & Yes \\ \hline
\end{tabular}
\end{table*}

\emph{Case II: $T > N$.} 
Define a vector $\Phi = (\phi_1, \phi_2, \cdots, \phi_{N^2})$
where $\phi_k$ is the number of collision-free slots in the $k$-th frame of the super frame.
We call $\Phi$ the collision-free-slots distribution of a sequence scheme.
We let $\Phi^{\textsf{TDMA}}=(\phi_1^{\textsf{TDMA}}, \phi_2^{\textsf{TDMA}}, \cdots, \phi_{N^2}^{\textsf{TDMA}})$
and $\Phi^{\textsf{GF}}=(\phi_1^{\textsf{GF}}, \phi_2^{\textsf{GF}}, \cdots, \phi_{N^2}^{\textsf{GF}})$
to respectively denote the collision-free-slots distributions of TDMA and the GF sequence scheme.
Note that we have $\sum_{i=1}^{N^2} \phi_i^{\textsf{TDMA}}  = NT$, and $\sum_{i=1}^{N^2} \phi_i^{\textsf{GF}} \le NT$.

We define a function $h(\Phi) = \sum_{k=1}^{N^2} \left[ 1- (1-p_i)^{\phi_{k}} \right]$,
which is the expected number of delivered packets before expiration in the super frame.
Note that $h(\Phi)$ increases as $\phi_k$ increases. Thus, for any vector $\Phi^{\textsf{GF}}$ with $\sum_{i=1}^{N^2} \phi_i^{\textsf{GF}} \le NT$,
we can always find another vector $\tilde{\Phi} = (\tilde{\phi}_1, \tilde{\phi}_2, \cdots, \tilde{\phi}_{N^2})$
with $\phi_i^{\textsf{GF}}  \le \tilde{\phi}_i \le T (\forall i=1,2,\cdots, N^2)$ and $\sum_{i=1}^{N^2} \tilde{\phi}_i = NT$ such that $h(\Phi^{\textsf{GF}} ) \le h(\tilde{\Phi})$.
Thus, it suffices to prove $h(\Phi^{\textsf{TDMA}}) \ge h(\Phi^{\textsf{GF}})$
when $\sum_{i=1}^{N^2} \phi_i^{\textsf{GF}}  = NT$. We will prove it next.

For vector $\Phi$, we further sort it in ascending order and get another vector
$\Phi_{\textsf{sorted}} = \left(\phi_{(1)},\phi_{(2)},\cdots, \phi_{(N^2)} \right)$
where $\phi_{(k)}$ is the $k$-th smallest element in vector $\Phi$.
We call $\Phi_{\textsf{sorted}}$ the sorted collision-free-slots distribution of a sequence scheme.
Clearly $h(\Phi)=h(\Phi_{\textsf{sorted}})$.

In this case of $T>N$,  under the uniformly-distributed TDMA sequence, similar to the analysis in the proof of Case II in Theorem~\ref{thm:theory-R-i-L-T},
each of the $N^2$ frame in this super frame is either a \emph{Type-1 frame}
where there are  $ \left \lceil T/N \right \rceil  $ collision-free slots or a \emph{Type-2 frame}
where there are  $ \left \lfloor T/N \right \rfloor  $ collision-free slots.\footnote{ It is straightforward to prove case II when $T \bmod N = 0$. Thus, in the rest of this case, we assume that
$T \bmod N \neq 0$. }
In addition,
there are $\alpha = N(T \bmod N)$ Type-1 frames and $\beta = N^2 - \alpha$ Type-2 frames in this super frame.
Then the sorted collision-free-slots distribution of TDMA is
\be
 \resizebox{.892\linewidth}{!}{$
\Phi^{\textsf{TDMA}}_{\textsf{sorted}}
= \left( \underbrace{\left \lfloor T/N \right \rfloor,\cdots, \left \lfloor T/N \right \rfloor}_{\beta \text{ times}},
\underbrace{\left \lceil T/N \right \rceil, \cdots, \left \lceil T/N \right \rceil}_{\alpha \text{ times}} \right).$}
\label{equ:phi-TDMA-sorted}
\ee

The sorted collision-free-slots distribution of the GF sequence scheme is
$
\Phi^{\textsf{GF}}_{\textsf{sorted}} = \left(\phi^{\textsf{GF}}_{(1)},\phi^{\textsf{GF}}_{(2)},\cdots, \phi^{\textsf{GF}}_{(N^2)} \right)$ with $\sum_{i=1}^{N^2} \phi^{\textsf{GF}}_{(i)}=NT$.
Next we only need to show that
\be
h(\Phi^{\textsf{TDMA}}) = h(\Phi^{\textsf{TDMA}}_{\textsf{sorted}}) \ge h(\Phi^{\textsf{GF}}) = h(\Phi^{\textsf{GF}}_{\textsf{sorted}}).
\label{equ:h-TDMA-larger-than-h-GF}
\ee

Toward that end, we need to use the following inequality,
\bee
& \left[1-(1-p_i)^{n_1} \right] + \left[1-(1-p_i)^{n_2} \right] \nnb \\
& \le  \left[1-(1-p_i)^{n_1+1} \right] + \left[1-(1-p_i)^{n_2-1} \right],
\label{equ:ineq-n-1-n-2-n-1+1-n-2-1}
\eee
for any nonnegative integers $n_1$ and $n_2$ satisfying $n_2-n_1 > 1$.
The proof of \eqref{equ:ineq-n-1-n-2-n-1+1-n-2-1} is straightforward and omitted here.

Now for $\Phi^{\textsf{GF}}_{\textsf{sorted}} = \left(\phi^{\textsf{GF}}_{(1)},\phi^{\textsf{GF}}_{(2)},\cdots, \phi^{\textsf{GF}}_{(N^2)} \right)$,
we do the following transformation.
If $\phi^{\textsf{GF}}_{(N^2)} - \phi^{\textsf{GF}}_{(1)} > 1$, i.e., the largest gap of $\Phi^{\textsf{GF}}_{\textsf{sorted}}$
is larger than 1, we then increase $\phi^{\textsf{GF}}_{(1)}$ by 1 and decrease $\phi^{\textsf{GF}}_{(N^2)}$ by 1 and sort the obtained vector again in ascending order
to get another new sorted vector
$\tilde{\Phi}^{\textsf{GF}}_{\textsf{sorted}} = \left(\tilde{\phi}^{\textsf{GF}}_{(1)},\tilde{\phi}^{\textsf{GF}}_{(2)},\cdots, \tilde{\phi}^{\textsf{GF}}_{(N^2)} \right)$.
Clearly, we have
\be
\tilde{\phi}^{\textsf{GF}}_{(1)}+\tilde{\phi}^{\textsf{GF}}_{(2)}+\cdots + \tilde{\phi}^{\textsf{GF}}_{(N^2)} = NT. \nnb
\ee
Based on \eqref{equ:ineq-n-1-n-2-n-1+1-n-2-1}, we further have
\be
h(\Phi^{\textsf{GF}}_{\textsf{sorted}}) \le h(\tilde{\Phi}^{\textsf{GF}}_{\textsf{sorted}}).
\label{equ:h-value-order-preseving}
\ee
Note that the largest gap of $\tilde{\Phi}^{\textsf{GF}}_{\textsf{sorted}}$ could be equal to or smaller
than the largest gap of ${\Phi}^{\textsf{GF}}_{\textsf{sorted}}$. We can keep doing such transformation until
the largest gap of the obtained vector is not greater than 1. The final obtained vector is denoted by
$
\bar{\Phi}^{\textsf{GF}}_{\textsf{sorted}} = \left(\bar{\phi}^{\textsf{GF}}_{(1)},\bar{\phi}^{\textsf{GF}}_{(2)},\cdots, \bar{\phi}^{\textsf{GF}}_{(N^2)} \right)$,
which satisfies
\be
\bar{\phi}^{\textsf{GF}}_{(1)}+\bar{\phi}^{\textsf{GF}}_{(2)}+\cdots + \bar{\phi}^{\textsf{GF}}_{(N^2)} = NT,
\label{equ:bar-phi-sum-NT}
\ee
and
\be
\bar{\phi}^{\textsf{GF}}_{(N^2)} - \bar{\phi}^{\textsf{GF}}_{(1)} \le 1.
\label{equ:bar-phi-gap-less-than-1}
\ee
In addition, \eqref{equ:h-value-order-preseving} yields to
\be
h(\Phi^{\textsf{GF}}_{\textsf{sorted}}) \le h(\bar{\Phi}^{\textsf{GF}}_{\textsf{sorted}}).
\label{equ:h-value-order-preseving-final}
\ee

To prove \eqref{equ:h-TDMA-larger-than-h-GF}, we next show that $\bar{\Phi}^{\textsf{GF}}_{\textsf{sorted}} = {\Phi}^{\textsf{TDMA}}_{\textsf{sorted}}$.

Note that \eqref{equ:bar-phi-sum-NT} indicates
$
N^2 \bar{\phi}^{\textsf{GF}}_{(N^2)} \ge NT \ge N^2  \bar{\phi}^{\textsf{GF}}_{(1)},
$
i.e., $\bar{\phi}^{\textsf{GF}}_{(1)} \le \frac{T}{N},
\bar{\phi}^{\textsf{GF}}_{(N^2)} \ge \frac{T}{N} $.
Then according to \eqref{equ:bar-phi-gap-less-than-1}, we must have
\be
\bar{\phi^{\textsf{GF}}}_{(1)} = \left \lfloor \frac{T}{N} \right \rfloor, \quad
\bar{\phi^{\textsf{GF}}}_{(N^2)} = \left \lceil \frac{T}{N} \right \rceil. \nnb
\ee
Thus, any entry in $\bar{\Phi}^{\textsf{GF}}_{\textsf{sorted}}$ is either
$ \left \lfloor \frac{T}{N} \right \rfloor$ or  $\left \lfloor \frac{T}{N} \right \rfloor$. Suppose that the number of
$\left \lceil \frac{T}{N} \right \rceil$ is $x$. From \eqref{equ:bar-phi-sum-NT}, we have
\be
x \left \lceil \frac{T}{N} \right \rceil + (N^2 - x) \left \lfloor \frac{T}{N} \right \rfloor = NT, \nnb
\ee
yielding to $x =  N(T \bmod N) = \alpha.$
Therefore, in vector $\bar{\Phi}^{\textsf{GF}}_{\textsf{sorted}}$, the number of
$ \left \lfloor \frac{T}{N} \right \rfloor$ is $N^2-x=N^2-\alpha = \beta$ and
the number of $ \left \lceil \frac{T}{N} \right \rceil$ is $x=\alpha$. Therefore, according to \eqref{equ:phi-TDMA-sorted}, we have
$\bar{\Phi}^{\textsf{GF}}_{\textsf{sorted}} = {\Phi}^{\textsf{TDMA}}_{\textsf{sorted}}.$
Therefore, \eqref{equ:h-value-order-preseving-final} indicates
\be
h(\Phi^{\textsf{GF}}_{\textsf{sorted}}) \le h(\bar{\Phi}^{\textsf{GF}}_{\textsf{sorted}}) = h(\bar{\Phi}^{\textsf{TDMA}}_{\textsf{sorted}}), \nnb
\ee
which proves \eqref{equ:h-TDMA-larger-than-h-GF}.

The proof is thus completed.

\subsection{Proof of Theorem~\ref{thm:compare-q-1-N-and-N}} \label{app:proof-of-thm-compare-q-1-N-and-N}
We prove this result by enumeration.
For different $N$, we list $q(1,N)$ and compare $q^2(1,N)$ with $N$ in Table~\ref{tab:q-1-N}.
As we can see, when $1 \le N \le 8$, we have $q^2(1,N) >  N$, i.e.,
the period of GF sequence is larger than the period of TDMA sequence.
When $N = 9$, we have $q^2(1,N) = N$, i.e.,
the period of GF sequence is equal to the period of TDMA sequence.
When $N \ge 10$, we consider the following different ranges.

\begin{itemize}
\item When $10 \le N \le 27$, we obtain that $q(1,N)=3$ where the corresponding $k=2$ in \eqref{equ:min-q-GF}. Thus, we have $q^2(1,N)=9 < 10 \le N$.
\item When $28 \le N \le 64$, we  obtain that $q(1,N)=4$ where the corresponding $k=2$ in \eqref{equ:min-q-GF}. Thus, we have $q^2(1,N)=16 < 28 \le N$.
\item When $N > 64$, we have $\lceil N^{\frac{1}{3}} \rceil \ge 4$ and $N^{\frac{1}{2}} > 2 N^{\frac{1}{3}}$. Thus according to Bertrand's postulate,
there exists  at least one prime number $q$ (which is of course a prime power) such that
\[
\lceil N^{\frac{1}{3}} \rceil < q < 2 \lceil N^{\frac{1}{3}} \rceil - 2.
\]
In addition, we have
\[
2 \lceil N^{\frac{1}{3}} \rceil - 2 < 2 N^{\frac{1}{3}} <  N^{\frac{1}{2}}.
\]
Thus, we can find a prime $q$ such that
\be
N^{\frac{1}{3}} \le \lceil N^{\frac{1}{3}} \rceil < q <  N^{\frac{1}{2}},
\ee
implying that
\[
q^3 \ge N, \quad q^2 < N.
\]
Thus when we set $k=2$ in \eqref{equ:min-q-GF}, this prime number $q$  satisfies \eqref{equ:min-q-GF} with $D=1$.
Since $q(1,N)$ is the minimal $q$ satisfying  \eqref{equ:min-q-GF} with $D=1$, we have
\be
q(1,N) \le q < \sqrt{N}.
\ee
Thus when $N \ge 64$, we have $q^2(1,N) < N$.
\end{itemize}

Thus, when $N \ge 10$, we have $q^2(1,N) < N$, i.e., the sequence period of the GF sequence scheme is shorter than that
of TDMA. According to Lemma~\ref{lem:R-1-R-2-decrease-with-L},
the average system timely throughput of TDMA is smaller than or equal to the lower bound
of the average system timely throughput of the GF sequence scheme, and of course smaller than the actual
average system timely throughput of the GF sequence scheme when $N \ge 10$.

The proof is thus completed.

\end{appendix}

\fi

\end{document}